\documentclass[12pt]{article}
\pdfoutput=1

\newlength{\abstractwidth}
\usepackage{amsmath}
\usepackage{amsfonts}
\usepackage{amssymb}
\usepackage{latexsym}

\usepackage{epsf}

\usepackage{color}
\usepackage{graphicx}
\usepackage{tikz}
\usepackage{dsfont}
\usepackage{hyperref}
\usepackage{braket}
\hypersetup{colorlinks=true, 
    linktoc=all,     
    linkcolor=black,  
    citecolor=blue
    }

\renewcommand{\thefootnote}{\fnsymbol{footnote}}
\renewcommand{\thanks}[1]{\footnote{#1}}
\newcommand{\starttext}{
\setcounter{footnote}{0}
\renewcommand{\thefootnote}{\arabic{footnote}}}

\newcommand{\bea}{\begin{eqnarray}}
\newcommand{\eea}{\end{eqnarray}}
\newcommand{\be}{\begin{eqnarray}}
\newcommand{\ee}{\end{eqnarray}}
\newcommand{\bma}{\begin{matrix}}
\newcommand{\ema}{\end{matrix}}
\newcommand{\<}{\langle}
\renewcommand{\>}{\rangle}

\def\cD{{\cal D}}

\def\cG{{\cal G}}

\def\cJ{{\cal J}}
\def\cK{{\cal K}}
\def\cL{{\cal L}}
\def\cM{{\cal M}}
\def\cN{{\cal N}}
\def\cO{{\cal O}}

\def\cQ{{\cal Q}}

\def\cT{{\cal T}}

\def\cZ{{\cal Z}}

\def\mS{\mathfrak{S}}

\def\ma{\mathfrak{a}}
\def\mb{\mathfrak{b}}

\def\RR{{\mathbb R}}
\def\NN{{\mathbb N}}
\def\CC{{\mathbb C}}

\def\Re{{\rm Re \,}}

\def\Tr{{\rm Tr}}
\def\det{{\rm det \,}}

\def\half{{1\over 2}}
\def\p{\partial}

\def\a{\alpha}
\def\b{\beta}

\def\eps{\epsilon}

\def\ep{\varepsilon}

\def\no{\nonumber}
\def\sm{\smallskip}

\def\W{\tilde W}

\def\Jb{\bar J}
\def\zb{\bar z}
\def\p{\partial}
\def\pz{\partial_z}
\def\pzb{\partial_{\bar z}}
\def\rt{\rightarrow}
\def\Jc{{\cal J}}
\def\Jct{{{\tilde{\cal J}}}}
\def\ve{\varepsilon}
\def\slt{$SL(2,\RR)$}

\begin{document}
\starttext
\setcounter{footnote}{0}

\begin{flushright}
\end{flushright}

\vskip 0.3in

\begin{center}

{\Large \bf Renormalization of gravitational Wilson lines\footnote{Research supported in part by  the National Science Foundation under grant PHY-16-19926.}}

\vskip 0.3in

{\large  Mert Be\c sken, Eric D'Hoker, Ashwin Hegde and Per Kraus}

\vskip 0.1in

{ \sl Mani L. Bhaumik Institute for Theoretical Physics}\\
{\sl Department of Physics and Astronomy }\\
{\sl University of California, Los Angeles, CA 90095, USA}

\vskip 0.05in

{\tt \small mbesken, dhoker, ashwin.hegde, pkraus@physics.ucla.edu}

\end{center}

\begin{abstract}
We continue the  study of the Wilson line representation of conformal blocks in two-dimensional conformal field theory; these have an alternative interpretation as gravitational Wilson lines in the context  of the AdS$_3$/CFT$_2$ correspondence.  The gravitational Wilson line involves a path-ordered exponential of the stress tensor, and its expectation value can be computed perturbatively in an expansion in inverse powers of the
central charge $c$. The short-distance singularities which occur in the associated stress tensor correlators   require systematic regularization and renormalization prescriptions, whose consistency with conformal Ward identities presents a subtle problem. The regularization used here combines dimensional regularization and analytic continuation. Representation theoretic arguments, based on $SL(2,\RR)$ current algebra, predict an exact result for the Wilson line anomalous dimension and, by building on previous work, we verify that the perturbative calculations using our regularization and renormalization prescriptions reproduce the exact result  to order  $1/c^3$ included.  We also discuss a related, but somewhat simpler, Wilson line in Wess-Zumino-Witten models that yields current algebra conformal blocks, and we emphasize the
distinction between Wilson lines constructed out of non-holomorphic and purely holomorphic currents.
\end{abstract}

\newpage
\tableofcontents

\baselineskip=15pt
\setcounter{equation}{0}
\setcounter{footnote}{0}



\setcounter{tocdepth}{1}


\section{Introduction}
\setcounter{equation}{0}
\label{sec:1}

Wilson lines and Wilson loops are obtained by the path-ordered exponential integral of a connection respectively along an open interval and a closed contour. In gauge theory, the connection is the canonical gauge field and the resulting Wilson loop operator is a gauge-invariant observable with applications to elucidating the phases of gauge theory and beyond. A different type of Wilson line operator has recently found use in two-dimensional conformal field theory; in this case the connection is a composite field involving the stress tensor of the CFT. What this object yields is a conformal block associated with a pair of primary operators, one at each endpoint of the Wilson line.   Actually, the two types of Wilson lines just mentioned are closely related objects if viewed in the context of the AdS/CFT correspondence: the CFT Wilson line is the boundary image of a bulk  Wilson line, and for this reason we often refer to it as a gravitational Wilson line, although it exists as an object in CFT independent of the AdS/CFT correspondence.   In this paper we continue the study of these Wilson lines, focussing in particular on their status as well defined quantum mechanical operators.   Their renormalization poses a rather subtle and nonstandard problem which we aim to understand better.

\sm

The general connection between Wilson lines in three dimensions and conformal field theory in two dimensions arose in \cite{Witten:1988hf}, and the relation to the Virasoro algebra appeared in \cite{Verlinde:1989ua}.  More recently, Wilson lines arose in the context of the AdS$_3$/CFT$_2$ correspondence, first as a tool for computing entanglement entropy in higher spin theories \cite{deBoer:2013vca,Ammon:2013hba}, and then in the more general context of computing conformal blocks \cite{Bhatta:2016hpz,Besken:2016ooo}.  Quantum aspects of these Wilson lines have been studied in \cite{Fitzpatrick:2016mtp,Besken:2017fsj,Anand:2017dav,Hikida:2017ehf,Hikida:2018dxe,
Hikida:2018eih}.  The related representation of CFT conformal blocks and OPE structures in terms of AdS appeared in \cite{Hijano:2015zsa} and in \cite{Czech:2016xec}.
We also note that the notion of integrating the stress tensor over a contour arises in the context of the averaged null energy condition (proven in flat space in  \cite{Faulkner:2016mzt,Hartman:2016lgu}), and the related notion of a ``length operator" discussed in \cite{Afkhami-Jeddi:2017rmx} has connections to the Wilson line discussed here.

\sm

More motivation and details on the form of the Wilson line will be given in the next section, but for now it suffices to write,
\bea
\label{W1}
W[z_2,z_1]= \langle j, -j|P \exp \left \{\int_{z_1}^{z_2} dz \Big (L_1 +\frac{6}{ c}T(z)L_{-1} \Big ) \right \} |j,j\rangle.
\eea
Except for the non-holomorphic Wilson line discussed in section \ref{sec:nonhowil} our formulas refer to a chiral half of a CFT and $z$ denotes the corresponding holomorphic coordinate on the plane. The shape of the integration contour from $z_1$ to $z_2$ used to define the Wilson line is inconsequential,  except when we introduce a regulator and break conformal invariance, and then it is taken to be along the real line. $L_0$ and $L_{\pm 1}$ are generators of the Lie algebra of \slt, with the states $|j,\pm j\rangle$ being highest/lowest weight states of a spin $j$ representation, corresponding to a primary of dimension $h=h(j,c)$, as will be discussed in more detail below.  $T(z)$ is the stress tensor operator, such that $W[z_2,z_1]$ is supposed to represent the  Virasoro vacuum OPE block corresponding to the bi-local $O(z_2)O(z_1)$, where $O(z)$ is a primary operator of dimension $h(j,c)$.  That is, $W[z_2,z_1]$ captures all terms in the $O(z_2)O(z_1)$ OPE involving only stress tensors.

\sm

The Virasoro vacuum block is a rich object, capturing as it does the effect of an arbitrary number of stress tensors.   Phrased in terms of AdS, it encodes the gravitational interaction~\cite{Hartman:2013mia}.  The Wilson line provides an expression for the Virasoro vacuum block in a form admitting a convenient $1/c$ expansion, which in the bulk corresponds to an expansion in Newton's constant.  Our goal here is to understand this perturbative expansion; once that is under control one can contemplate using the Wilson line to study non-perturbative effects as well.

\sm

The Wilson line as defined in (\ref{W1}) is a singular object due to the appearance of stress tensors at coincident points, and thus requires regularization and renormalization \cite{Fitzpatrick:2016mtp,Besken:2017fsj,Hikida:2017ehf}.  Here we  adopt a type of dimensional regularization \cite{Hikida:2017ehf}, in which the stress tensor is taken to have dimension $2-\ve$. Renormalization of the  Wilson line then requires an overall multiplicative renormalization by a factor $N(\ve)$, as well as a vertex renormalization factor $\alpha(\ve)$ multiplying $T(z)$, where both $N(\ep)$ and $\alpha(\ep)$ depend on the regulator $\ep$ as well as on $c$ and $j$.  This regularization scheme breaks conformal invariance at intermediate stages, and from the point of view of diagrammatics it is highly nontrivial that conformal invariance is restored upon renormalization.

\sm

The most basic quantity to consider is the Wilson line expectation value itself; given what we have said, this should take the form of a conformal two-point function,
\bea
\langle W[z_2,z_1]\rangle \sim |z_2-z_1|^{-2h(j,c)}.
\eea
At lowest order in the $1/c$ expansion one finds $h(j,c)=-j$, but this value receives corrections order by order in an expansion in powers of $1/c$.  There is in fact an expectation for the exact answer based on general conformal field theory considerations.  The Wilson line, as we have defined it, is based on a representation of \slt\ but once the stress tensors are included it describes an object in Virasoro representation theory.   Hamiltonian reduction supplies a procedure for constructing a representation of the Virasoro algebra by imposing a constraint on a corresponding representation of \slt\ current algebra.  This procedure has an analog in bulk gravity, where the constraints are precisely those that correspond to imposing asymptotically AdS boundary conditions.     The resulting relation between the \slt\ spin $j$ and the Virasoro dimension $h(j,c)$ is given by, (see e.g. \cite{Bershadsky:1989mf}),
\bea\label{aaa} h(j,c)=-j+\frac{m+1}{ m} j(j+1)~,
\hskip 0.8in
  c=1-{6\over m(m+1)}~.
  \eea
Expanding $h(j,c)$ in powers of $1/c$ the first few contributions are given by
\bea
\label{aab}
h(j,c)=-j-{6\over c}j(j+1)-{78\over c^2}j(j+1)-{1230\over c^3}j(j+1) + {\cal O}(c^{-4})~.
\eea
and provide a prediction for the perturbative expansion of the Wilson line expectation value.  One of the main results of this paper is to verify, by explicit calculation,  that the procedure of dimensional regularization and renormalization via the inclusion of the factors $N(\ep)$ and $\alpha (\ep)$, does indeed reproduce the dimension formula (\ref{aab})  to the order indicated, thereby extending previous results \cite{Besken:2017fsj,Hikida:2017ehf}.

\sm

It is also useful to give a bulk gravity perspective on the result (\ref{aab}) in terms of gravitational self-energy.   If we take the classical point particle limit, $c,j \rt \infty$, with $j/c$ fixed we can write the result as $m=m_0-2Gm_0^2$.  To obtain this we used the Brown-Henneaux formula $c={3\ell\over 2G}$, the relation between the mass of a particle in AdS and the corresponding conformal dimension $m\ell =2h$, and similarly $m_0 \ell = 2h_0=-2j$.  The relation between $m$ and $m_0$ is the same as that obtained from considering the classical gravitational self-energy of a point particle in AdS \cite{Besken:2017fsj}.  The general formula (\ref{aab}) can thus be thought of as supplying quantum corrections to this result. This is interesting, because the gravitational self-energy is typically ill-defined, or rather sensitive to  unknown UV physics, but the situation in three dimensions appears to be under better control.

\subsection{Organization}

We now summarize the remainder of this paper.  In section \ref{Wilsec} we review the logic behind the construction of the gravitational Wilson line.  In section \ref{wzw} we discuss the analog of the gravitational Wilson line for a level $k$ current algebra with conserved current $J^a(z)$ given by the Wilson line $ P \exp {1\over k}\int  J^a T^a $. Here $T^a$ denote the generators of the relevant Lie algebra, and we denote by $G$ the corresponding Lie group. This object yields the current algebra vacuum OPE block for a bi-local primary operator.  Its evaluation poses a similar, but somewhat simpler, renormalization problem as compared to the stress tensor case.  In this case the $1/k$ expansion should yield the standard expression for the scaling dimension $h$ of a current algebra primary in terms of quadratic Casimirs, $h = C_2(r)/(2k+C_2(G))$.   In order to better understand the origin of the Wilson line, we study an alternative construction starting from the WZW model. We consider the bi-local operator $g^{-1}(x_2) g(x_1)$ constructed from the basic WZW primary $g(x)$ which lives on the group manifold $G$. This can be written identically in terms of a Wilson line for a  non-conserved current, $\Jc_\mu =-k g^{-1} \p_\mu g$, and admits a relatively straightforward perturbative expansion using standard dimensional regularization (modulo subtleties associated with the appearance of epsilon tensors).  What is not manifest in this approach is why this operator holomorphically factorizes.

\sm

In section \ref{sec:5} we turn to the gravitational Wilson line. We describe the systematics of the renormalization procedure and compute the expectation value of the Wilson line with zero and one additional stress tensor insertions through order $1/c^3$.   Consistency of these two computations uniquely fixes all renormalization constants and yields an unambiguous answer for the anomalous dimension, which indeed reproduces the expansion (\ref{aab}).  In section \ref{sec:6} we discuss an alternative regularization procedure.  Rather than modifying the dimension of the stress tensor we adopt another method for softening the short distance singularities arising from collisions of stress tensors.  This approach also involves a dimensionless regulator $\ve$ and {\em a priori} seems just as sensible as the prior scheme.  However, our explicit computations reveal that conformal invariance is not recovered in this scheme.  This serves to highlight the subtleties involved in renormalizing the Wilson line.     We close the paper with some comments in section \ref{sec:7}.   Various technical results appear in appendices.

\section{The gravitational Wilson line operator}
\setcounter{equation}{0}
\label{Wilsec}

Consider a primary operator $O(z, \bar z)$ in a two-dimensional CFT. As most of our considerations involve one chiral half of the CFT, we henceforth write $O(z)$.  Under a conformal transformation, $z'=f(z)$, the bi-local operator $O(z_2)O(z_1)$ transforms as
\bea\label{aa}
O(z_2')O(z_1') =  \big(f'(z_2)f'(z_1)\big)^{-h}O(z_2)O(z_1)~,
\eea
which identifies the scaling dimension $h$ of $O$.

\subsection{Wilson line covariant under global conformal transformations}

We first discuss how to write down a Wilson line whose transformation is given by (\ref{aa})  under global conformal transformations, $f(z)= (az+b)/(cz+d)$, which describe an \slt\ subgroup of the full Virasoro symmetry.  To this end, let $(L_{-1},L_0,L_1)$ be \slt\ generators obeying $[L_m,L_n]=(m-n)L_{m+n}$.   We then consider the matrix element
\bea\label{ab} W[z_2,z_1]=\langle h; {\rm out}|  P \exp \left \{  \int_{z_1}^{z_2} dz L_1  \right \} |h;{\rm in}\rangle~,\eea
for suitable in and out states to be defined momentarily.

\sm

To see how to implement the conformal transformation, consider the more general path ordered exponential $P \exp\int_{z_1}^{z_2}a(z)$, where the connection $a(z)=a_z(z)dz$ takes values in the Lie algebra of \slt.  Under the action of an arbitrary group element $U(z) \in$ \slt, the connection transforms by $U^{-1}(z) a(z)U(z)-U^{-1}(z)dU(z)= a_U(z)$ while the Wilson line transforms by
\bea\label{aba}
U^{-1}(z_2) P \exp \left \{ \int_{z_1}^{z_2}a(z)\right \} U(z_1) = P \exp \int_{z_1}^{z_2} a_U(z)~.
\eea
In the present case, $a(z)=L_1 dz$. The following transformation leaves $a(z)$ invariant, i.e. $a_U(z)=a(z)$, and hence represents a global conformal transformation
\bea\label{ac}
U(z) = e^{\lambda_1(z)L_1} e^{\lambda_0(z)L_0} e^{\lambda_{-1}(z) L_{-1} }
\eea
with
\bea\label{ad}
\lambda_1 = z-f(z)~,\qquad \lambda_0(z) =- \ln (f'(z))~,\qquad \lambda_{-1}(z) = -{f''(z)\over 2f'(z)}~,
\eea
and $f(z)=(az+b)/(cz+d)$ as above.   Together with (\ref{aba}) we then have
\bea
\label{ae}
W[z_2,z_1]=\langle h; {\rm out}| e^{-\lambda_{-1}(z_2)L_{-1}} e^{ \ln[f'(z_2)]L_0}
P \exp \Big \{ \int_{z'_1}^{z'_2} dz L_1  \Big \}
e^{-\ln[f'(z_1)]L_0} e^{\lambda_{-1}(z_1)L_{-1}}|h;{\rm in}\rangle,~
\eea
again with $z'=f(z)$.  We now observe that if the states are taken to obey
\bea\label{af}
&&L_{-1}|h;{\rm in}\rangle =0~,\quad L_0 |h;{\rm in}\rangle =-h|h;{\rm in}\rangle \cr
&& L_{1}|h;{\rm out}\rangle =0~,\quad L_0 |h;{\rm out}\rangle =h|h;{\rm out}\rangle~,
\eea
then we obtain the desired transformation law
\bea\label{ag}
W[z'_2,z'_1] =  \big(f'(z_2)f'(z_1)\big)^{-h}W[z_2,z_1]~.
\eea
It will be convenient to write $h=-j$, since if $2j$ is a non-negative integer the $L_n$ can be taken to be a finite dimensional matrix representation of \slt.  One can then carry out computations for such $j$ and at the end set $j=-h$ for $h\geq 0$.   This is just a computational shortcut, and the same results are obtained by working with representations with $h \geq 0$ throughout. A convenient representation for $h\geq 0$ is discussed in appendix \ref{sec:AA}.

\sm

With this in mind, our Wilson line is at this stage written as
\bea
\label{WLL}
W[z_2,z_1]= \langle j, -j|  P \exp \Big \{ \int_{z_1}^{z_2} dz L_1  \Big \} |j,j\rangle
\eea
with $L_0|\pm j \rangle = \pm |\pm j\rangle$ and $L_{\mp 1}|\pm j\rangle =0$.  The matrix element is readily evaluated  using the fact that $L_1$ lowers the $L_0$ eigenvalue by one, and we have $W[z_2,z_1] \sim z^{2j} = z^{-2h}$.   The Wilson line (\ref{WLL})  thus gives the coefficient of the identity operator in the OPE expansion of the two primaries:  $O(z_2)O(z_1) \sim W[z_2,z_1]+ $(other operators).

\sm

The Wilson line (\ref{WLL}) emerges naturally in the AdS/CFT correspondence when we describe gravity in the bulk in the Chern-Simons formulation.  The AdS metric $ds^2 = d\rho^2 + e^{2\rho}dzd\zb$ is represented by the pair of connections $A=e^\rho L_1 dz +L_0 d\rho$ and $\overline{A} = e^{\rho}L_{-1}d\zb  - L_0 d\rho$. See, e.g. \cite{Campoleoni:2010zq}. The Wilson line in the Chern-Simons theory  $W[z_2,z_1]=\langle j,-j| P\exp \big \{ \int_{z_1}^{z_2} A\big \} |j,j\rangle$ reduces to the Wilson line (\ref{WLL}) upon substituting for  $A$ with the reduced connection $a=L_1 dz$, as the $\rho$ dependence can be gauged away.

\sm

This ``global Wilson line" of (\ref{WLL}) forms the basis of a convenient description of arbitrary global  (i.e \slt) conformal blocks.  Rather than a single Wilson line, one considers a network with trivalent vertices.   Each vertex is represented by a singlet state in the tensor product of the three representations that enter the vertex.   The space of conformal blocks is in one-to-one correspondence with the space of such networks;  see \cite{Bhatta:2016hpz,Besken:2016ooo}.

\subsection{Wilson line covariant under local conformal transformations}

The main point of the preceding subsection was to motivate the form of the Wilson line that incorporates the stress tensor.  It should yield the Virasoro OPE block, which is to say that it should capture all contributions to the $O(z_2)O(z_1)$ OPE involving only stress tensors.  One way to motivate the proposal is to repeat the analysis that led to (\ref{ag}) but now for an arbitrary local conformal transformation $z'=f(z)$.   In this case, $a(z)$ cannot be left invariant, but must transform as follows
\bea
\label{ah}
a_U(z)= \left ( L_1 +{6\over c}T(z)L_{-1} \right ) dz
\eea
with $T(z)$ given in terms of $f(z)$ by
\bea\label{ai} T(z) =   {c\over 12} \, S_f(z)~,
\qquad S_f(z) = {f'''(z)\over f'(z)}-{3\over 2} \left( f''(z)\over f'(z)\right)^2~,
\eea
where $S_f(z)$ is the Schwarzian derivative.   We then obtain
\bea
\label{aj}
\langle j,-j| P \exp \Big \{ \int_{z_1}^{z_2} dz \Big ( L_1 +{6\over c}T(z)L_{-1} \Big ) \Big \} |j,j\rangle =  { [f'(z_2) f'(z_1)]^h \over [ f(z_2)-f(z_1)]^{2h}}~,
\eea
where we again have $h=-j$.   In this expression $T(z)$ is the classical function given in (\ref{ai}), not the stress tensor operator.  However, this result naturally suggests an expression for the Virasoro vacuum OPE block as  the gravitational Wilson line $W[z_2, z_1]$ given by,
\bea
\label{ak}
W[z_2,z_1]
\equiv \langle j,-j| P \exp \Big\{\int_{z_1}^{z_2} dz \Big ( L_1 +{6\over c}T(z)L_{-1} \Big ) \Big\}|j,j\rangle~,
\eea
where now $T(z)$ is the stress tensor operator.  In particular, suppose we take the expectation value of $W$  in a CFT state with a classical stress tensor expectation value in the large $c$ limit.   Such a stress tensor can be generated from $T(z)=0$ by some conformal transformation $z'=f(z)$.  The Wilson line expectation value should then be equal to the primary two-point function transformed by $f(z)$, and this is precisely what (\ref{aj}) says.  At the level of correlation functions, the statement that $W[z_2,z_1]$ is the Virasoro vacuum block is the statement that it equals $O(z_2)O(z_1)$ inside any correlation function involving just stress tensors,
\bea
\label{al}
\langle O(z_2)O(z_1) T(z_3)\ldots T(z_n)\rangle = \langle W[z_2,z_1] T(z_3)\ldots T(z_n)\rangle~. \eea
See \cite{Fitzpatrick:2016mtp} for more discussion and tests of this proposal.

This Wilson line also arises naturally from the bulk Chern-Simons description.  The most general asymptotically AdS$_3$ solution of Einstein's equations corresponds to the connections
\bea
A& = & (e^\rho L_1 +{6\over c} e^{-\rho}T(z)L_{-1})dz+L_0 d\rho
\no \\
\overline{A} & = & (e^\rho L_{-1} +{6\over c} e^{-\rho}\overline{T}(\zb)L_{1})dz-L_0 d\rho
\eea
where the holographic dictionary identifies $T(z)$ and $\overline{T}(\zb)$ as the components of the dual CFT stress tensor (e.g. \cite{Campoleoni:2010zq}).   The Wilson line therefore corresponds to  $P\exp\int a$ where $a= (L_1 +{6\over c}T(z)L_{-1})dz$ is the reduced connection.  In the quantum theory we should integrate over all asymptotically AdS connections weighted by the Chern-Simons action.  On general grounds, this should have the effect of replacing any string of stress tensors by their vacuum expectation value, and this is precisely what was meant above in saying that $T(z)$ appears in the Wilson line as an operator.


\section{Current algebra Wilson lines in the WZW model}
\setcounter{equation}{0}
\label{wzw}

Just as the gravitational Wilson line defined in terms of the stress tensor encodes conformal blocks of the Virasoro algebra, we can define a Wilson line built out of a spin-1 current that encodes current algebra conformal blocks.   The current algebra Wilson line is a somewhat simpler object, and we also have the useful Lagrangian realization of current algebra in terms of the WZW model.  In this section we explore this current algebra Wilson line from several complementary points of view. We first define a holomorphic Wilson line that is the direct spin-1 analog of our gravitational Wilson line and discuss its renormalization. We then turn to a non-holomorphic Wilson line, defined by a simple rewriting of a bi-local primary operator.  Its renormalizaton proceeds somewhat differently, but we show that  the anomalous dimensions of the two Wilson lines agree.  We finally make some comments about the connection between these two constructions.

\subsection{WZW model and current algebra}

We first review some background material; see, e.g. \cite{DiFrancesco:1997nk}.
The action of the WZW model is
\bea
\label{pa}
S[g] = {k\over 4\pi} \int_\Sigma\! d^2 x \sqrt{\gamma }\gamma ^{\mu \nu} \,
\Tr'(\p_\mu g^{-1} \p_\nu g)  +{ik \over 6\pi} \int_{\Gamma} \Tr'(\omega)^3
\eea
where the theory lives on the Riemann surface $\Sigma$ with local coordinates $x^\mu$, metric $\gamma_{\mu\nu}$, and where $\gamma = \det (\gamma_{\mu \nu})$. The surface $\Sigma$ is the boundary of a three-manifold $\Gamma$. The field $g(x)$ takes values in a compact Lie group $G$, and the one-form $\omega = g^{-1}dg$ takes values in the Lie algebra $\cG$ of $G$, in an arbitrary finite-dimensional irreducible representation $r$.  Denoting the structure constants of $\cG$ by $f^{abc}$ and a basis of Hermitian generators of $\cG$ in the representation $r$ of $\cG$ by $T^a$ with $a,b,c=1,\cdots , \dim \cG$, the structure relations are given by $[T^a,T^b]=\sum_c if_{abc}T^c$, and  we use the normalization for the trace in the representation $r$ by $\Tr ' (T^a T^b) = \half \delta ^{ab}$. With these normalization conventions, the level $k$ is quantized such that $2k$ is an integer.    We denote by $C_2(r)$ the value of the quadratic Casimir operator $C_2= \sum_a T^a T^a$ in representation~$r$.     For example, for $G=SU(N)$ and $r$  the defining representation we have $C_2(r)= (N^2-1)/(2N)$, while in the adjoint representation $r=G$ we have $C_2(G)=N$.

\sm

Invariance of $S[g]$ under global transformations $g(x) \to g_L \, g(x) \, g_R^{-1} $ with $(g_L, g_R) \in G\times G$ implies the existence of two independent conserved currents, which take the form
\bea
\label{pb}
 J^{\mu}  =  -{k\over 2}   (\gamma^{\mu\nu}-i\eps^{\mu\nu})g^{-1}\p_\nu g~,
 \hskip 1in
 \Jb^{\mu} =  -{k\over 2}   (\gamma ^{\mu\nu}+i\eps^{\mu\nu})\p_\nu g \, g^{-1} ~.
\eea
In a system of local complex coordinates $x^\mu = (z, \bar z)$, the metric on $\Sigma$  takes the form  $\gamma_{\mu \nu} dx^\mu dx^\nu = dzd\zb$ and the non-vanishing components of $\gamma_{\mu \nu}$ and of the anti-symmetric tensor $\eps_{\mu \nu}$ are given by $\gamma_{z \bar z} = \gamma _{\bar z z}= \half$ and $\eps_{z\zb}=- \eps_{\bar z z}={i\over 2}$. In terms of these complex coordinates,  the expressions for the currents simplify as follows
\begin{align}
\label{pc}
J_z & =-k \, g^{-1} \p_z g~, & J_{\zb} & =0~,
\no \\
\Jb_{\zb} & = -k \, \p_{\zb} g \, g^{-1}~, &  \Jb_z & =0~,
\end{align}
and obey $\pzb J_z = \pz \Jb_{\zb}=0$.  In view of these relations, $J_z$ and $\Jb _{\zb}$ are respectively referred to as the holomorphic and anti-holomorphic currents of the WZW theory, properties which will be reflected in the notation of their coordinate dependence  $J_z(z) $ and $\Jb _{\zb} (\zb)$.  The holomorphic currents $J_z(z) = \sum_a J^a_z(z)T^a$ obey the OPE
\bea\label{pcs} J^a_z(z) J^b_z(0) \sim {k\delta_{ab}\over z^2}+ \sum_c if_{abc} {J^c_z(0)\over z}~,
\eea
and similarly for the anti-holomorphic currents.

\sm

Primary operators of the WZW theory are group elements $g(x)$ taken in some representation $r$. They have conformal weight $(h,h)$ where the dimension $h$ is given by,
\bea\label{pd}
h = {C_2(r)\over 2k+C_2(G) }~.
\eea
The basic two-point function is $\langle g^{-1}(x_2) g(x_1)\rangle$, which is proportional to the identity matrix by virtue of the $G\times G$ global symmetry \footnote{In what follows we do not distinguish between a matrix proportional to the identity and one of its diagonal elements, including in the cases of the holomorphic and WZW Wilson lines.}.  Expanded in powers of $1/k$ we have
\bea
\label{pe} \langle g^{-1}(x_2)g(x_1)\rangle &\sim& (x_{21}^2)^{-2h}
\\
& \sim &1-{C_2(r)\over k} \ln(x_{21}^2)+{C_2(r)^2\over 2k^2} \big(\ln(x_{21}^2)\big)^2 + {C_2(r)C_2(G) \over 2k^2} \ln(x_{21}^2)  + \cO(k^{-3})
\no
\eea
up to an overall multiplicative  factor. Here we use the notation $x_{21} ^\mu = x^\mu _2 - x ^\mu _1$.

\sm

In perturbation theory in powers of  $1/k$  the scaling dimension $h$ is extracted from the correlator by computing Feynman diagrams.   The algebraic approach to the WZW model yields the full result (\ref{pd}), such that the perturbative series simply amounts to the shift $2k \rt 2k+ C_2(G)$.  In diagrammatic terms it is not at all obvious how we just get this simple shift.  However, agreement is expected,  since we  have good reason to believe that the path integral and algebraic definitions of the WZW theory describe one and the same theory.
Examples of perturbation theory computations in WZW include \cite{Bos:1987fb,deWit:1993qv}.  However, we are not aware of any prior computation of the anomalous dimension of primary operators in perturbation theory.

\subsection{Holomorphic Wilson line}

Given the holomorphic current $J(z)\equiv J_z(z)$ a natural object to consider is the Wilson line operator $P \exp{1\over k} \int_{z_1}^{z_2} dz J(z)$, where $P$ denotes path ordering along the contour from $z_1$ to $z_2$.   The basic claim is that, up to renormalization, this operator gives the current algebra vacuum OPE block.   That is, consider the bi-local operator $g^{-1}(x_2)g(x_1)$, with $g$ taken in some irreducible representation $r$.   We can decompose the operator into irreducible representations of the current algebra.  The Wilson line then gives the representation containing only holomorphic currents.  An equivalent way of stating this is that the Wilson line should reproduce correlation functions with any number of holomorphic current insertions,
\bea
\label{qa}
\langle g^{-1}(x_2)g(x_1) J^{a_3}(z_3) \ldots  J^{a_n}(z_n)\rangle =
\cZ(\bar z_1, \bar z_2)
\langle P \exp\Big\{{1\over k} \int_{z_1}^{z_2} \!\!\! dz J(z) \Big\} J^{a_3}(z_3) \ldots  J^{a_n}(z_n)\rangle
\eea
where the factor $\cZ(\bar z_1, \bar z_2)$ is independent of $z_3, \cdots, z_n$ and depends anti-holomorphically on $x_1$ and $x_2$ through $\bar z_1 $ and $\bar z_2$ only.
At lowest order in $1/k$ this is easy to establish using the OPE of the currents. At higher orders we encounter divergences requiring renormalization.   In this section we wish to check this relation in a perturbative  expansion in powers of  $1/k$ in terms of suitably renormalized operators. Setting $z_1=0$ we consider the case of zero and one current insertions, and we wish to establish
\bea
\label{qb}
W(z) \equiv \lim_{\ep \to 0} \, \<W_\ep(z) \> &=& z^{-2h}~, \cr
\lim_{\ep \to 0} \, \langle J^a(x)  W_\ep(z) \rangle &=& z^{-2h}\left( {1\over x-z}-{1\over x}  \right)  T^a \eea
where we introduced the notation $W_\ep(z)$ for the regulated Wilson line operator and $W(z)$ for its renormalized vacuum expectation value.

Although the current algebra Wilson line  can be understood on its own terms, it is usefully thought of as existing due to the well-known equivalence between Chern-Simons theory on a three-dimensional manifold $M$  and the WZW theory on the boundary of $M$ \cite{Witten:1988hf,Elitzur:1989nr}.  The natural observables in Chern-Simons theory are Wilson lines $P\exp\int A$, and in the present case we consider an open Wilson line with endpoints on the boundary.   On account of the flatness of the connection, the precise shape of the Wilson line contour does not matter, only the location of its endpoints, hence the Wilson line is well-suited to represent the bi-local operator $g^{-1}(x_2)g(x_1)$ (or more precisely, its current OPE block).

To flesh this out a bit more, the boundary components of the Chern-Simons gauge field are mapped in the WZW model to the current and an external gauge field:  $(A_z, A_{\zb})  \leftrightarrow (J_z, A_{\zb})$.  In the Chern-Simons path integral we fix $A_{\zb}$ on the boundary but allow $A_z$ to fluctuate.   Such a  path integral is equal on the WZW side to a generating function for the current correlators,  $\langle \exp\int\!d^2z J^a_z A^z_{\zb}\rangle$. This is established by relating the Chern-Simons equations of motion to the current algebra Ward identity.   The same procedure can be carried out in the presence of a Wilson line.   The Chern-Simons gauge field now gets a source due to the Wilson line, which maps on the WZW side to the Ward identity for the current in the presence of a primary operator inserted at each endpoint.  This then leads to the equivalence (\ref{qb}) between current correlators computed in the presence of the bi-local $g^{-1}(x_2)g(x_1)$ and in the presence of the Wilson line $P \exp{1\over k}\int_{z_1}^{z_2} dz J$.      This discussion explains why such a relation exists, but it is purely formal, as it does not take into account UV divergences in the path integral.     Here, we are trying to establish that the relation holds in the full quantum theory.

\subsection{Lowest order calculations}

The regulated holomorphic Wilson line operator takes the form
\bea\label{qc}W_\ep(z)=  N(\ve) P \exp\Big\{{\alpha(\ve)\over k} \int_{0}^{z} dy J(y) \Big \}~,
\eea
where $\ve$ is  a UV regulator. Expanding out the exponential and taking the vacuum expectation value, we need to compute nested integrals of current correlators. All current correlators are obtained from the corresponding modification of the standard recursion relation, which is determined by OPE and holomorphy considerations,
\bea
\label{qe}
&&
\langle J^a(y) J^{a_n}(y_n) \ldots J^{a_1}(y_1)\rangle
\no \\ && \hskip 0.4in
= \sum_{i=1}^n \sum_b {i f_{a a_i b} \over (y-y_i)^{1} } \langle  J^{a_n}(y_n) \ldots J^{a_{i+1}}(y_{i+1})J^b(y_i) J^{a_{i-1}}(y_{i-1}) \ldots J^{a_{1}}(y_{1})\rangle
\no \\ && \hskip 0.4in
+ \sum_{i=1}^n {k\delta_{aa_i} \over (y-y_i)^{2}}\langle J^{a_n}(y_n) \ldots J^{a_{i+1}}(y_{i+1)} J^{a_{i-1}}(y_{i-1}) \ldots J^{a_1}(y_{a_1}) \rangle  
\eea
starting from $\braket{\mathds{1}}=1$ and $\braket{J^a(y)=0}$. Singularities arise from collisions of pairs of currents, as in \cite{Hikida:2017ehf}.    We implement a form of dimensional regularization in which we assign scaling dimension $1-\ve$ to the currents.  For example, the regulated two-point function is
\bea
\label{qd}
\langle J^a(y_1) J^b(y_2)\rangle  = {k\delta_{ab}\over (y_1-y_2)^{2-2\ve} }~.
\eea
Regulating correlators can be subtle since each term in the recursion relation \eqref{qe} doesn't scale as the full correlator should. For example the three point function is obtained as
\bea
\braket{J^a(y_1)J^b(y_2)J^c(y_3)} &=&
\sum _d {if_{abd}\over (y_1-y_2)}{k\delta_{dc}\over (y_2-y_3)^2}+
\sum _d {if_{acd}\over (y_1-y_3)}{k\delta_{bd}\over (y_2-y_3)^2}\no
\\&=&{ikf_{abc}\over (y_1-y_2)(y_1-y_3)(y_2-y_3)}~.
\eea
Our prescription then is to first compute the correlator and write it in a form where scaling of each coordinate is manifest. Then simply replace every instance of $(y_i-y_j)$ with $(y_i-y_j)^{1-\ve}$.

\sm

To illustrate the general procedure outlined above, we consider the Wilson line expectation value at order $1/k$,
\bea\label{gf} \langle W_\ep(z) \rangle &=&
N (\ep) \Big[ 1+{\alpha(\ep) ^2 \over k^2} \sum_{a,b}T^a T^b \int_0^{z} dy_1 \int_0^{y_1} dy_2 \langle J^a(y_1)J^b(y_2)\rangle + \ldots   \Big] \cr&=&
N(\ep) \Big[ 1+{\alpha(\ep) ^2 \over k} C_2(r)  \int_0^{z} dy_1 \int_0^{y_1} dy_2 {1\over (y_1-y_2)^{2-2\ve}}  + \ldots   \Big]  \cr &=&
N(\ep) \left [ 1-{\alpha(\ep) ^2 \over k} C_2(r) \Big( {1\over 2\ve} + \ln z +1 + \cO(\ve)\Big)  + \ldots   \right ]
\eea
At this order we can take $N(\ve) = 1 +{C_2(r) \over 2k\ve}$ and $\alpha (\ep) =1$.  This gives the expected result
\bea
\label{qg}
W(z)  \sim z^{-2h}+ \cO\left({1\over k^2}\right)~,\qquad
h = {C_2(r)\over 2k} + \cO\left({1\over k^2}\right)~.
\eea
We can similarly verify the Ward identity at lowest order, which corresponds to expanding the exponential to first order.  This gives
\bea\label{qh} \langle J^a(w)  W_\ep(z) \rangle &=&   T^a \int_{0}^{z} {dy\over (y-w)^{2-2\ve}} + \ldots \cr& = &   T^a  \left( {1\over w-z}-{1\over x}\right) + \ldots   \eea
which is the correct result at this order.

\sm

\subsection{Higher order computations}

We now make a few comments about the computation of the holomorphic Wilson line at higher orders in $1/k$.  We will be brief here, as the most significant technical details will be discussed later in the context of the gravitational Wilson line.

\sm

Since the correlation function of $n$ currents contains a maximal power $k^p$ with $p=\left \lfloor{{n\over 2}}\right \rfloor$, to obtain the Wilson line at order $1/k^n$ we need to expand the exponential to order $2n$.  The correlation function of up to $2n$ currents is obtained from the recursion relation (\ref{qe}).  The nested integrals can be evaluated by the methods discussed below.  Finiteness of the renormalized Wilson line as $\ve \rt 0$ only partially fixes the renormalization constants $N(\ve)$ and $\alpha(\ve)$ up to the given order in the $1/k$ expansion.  The unfixed part of $N(\ve)$ can be fixed by adopting a normalization convention, such as $\langle W_\ep(1) \rangle=1$. To fix   $\alpha$, which is needed to determine the scaling dimension, we need to demand that the Ward identity is satisfied.  Rather than the general Ward identity (\ref{qb}), various integrals greatly simplify if we place the current at infinity, using the usual formula obtained from $z\rt 1/z$:  $J^a_\infty \equiv - \lim_{z\rt \infty} \, z^2 J^a(z)$.    So this amount to imposing
\bea
\label{qi}
\lim_{\ep \rt 0} \, \langle J^a_\infty W_\ep(z) \rangle = T^a z W(z) ~.\eea
 We carry this out order by order in $1/k$, fixing the constants $N$ and $\alpha$ up to that order as we go.  These considerations completely fix the terms in the $\ve$ expansion that contribute to the finite parts of the correlators as $\ve \rt 0$.  The program is in fact highly overconstrained, since just from counting terms there is no guarantee that constants $N$ and $\alpha$ can be found that satisfy these criteria.   It is furthermore not guaranteed that the Wilson line correlator will be a pure power law.  Nevertheless, explicit computations demonstrate that all these conditions are indeed satisfied, at least to third order in the $1/k$ expansion.

\sm

As an example, consider the Wilson line \eqref{qc} expanded to order $1/k^2$. Focussing only on the term that involves three current insertions, we have
\bea
\braket{W_\ve(z)}
&\sim&
{N\alpha^3\over k^3}\sum_{a,b,c}T^aT^bT^c\int_0^zdy_1\int_0^{y_1}dy_2\int_0^{y_2}dy_3\braket{J^a(y_1)J^b(y_2)J^c(y_3)}~,
\\
&=&{N\alpha^3\over k^2}\sum_{a,b,c}if_{abc}T^aT^bT^c
\int_0^z \!\! dy_1 \int_0^{y_1} \!\! dy_2\int_0^{y_2} \!\! dy_3
{1\over (y_1-y_2)^{1-\ve}(y_1-y_3)^{1-\ve}(y_2-y_3)^{1-\ve}}~.\no
\eea
The integral is discussed in detail in the gravitational case and we will skip its derivation  here. The Lie algebra factor multiplying the integral is easily computed as
\bea
if_{abc}T^aT^bT^c &=&\tfrac{1}{2}\sum_{a,b,c,d}if_{abc}[T^a,T^b]T^dT^c~,\no
\\&=&-\tfrac{1}{2}\sum_{a,b,c,d}f_{abc}f_{abd}T^dT^c~,\no
\\&=&-\tfrac{1}{2}C_2(G)C_2(r)~,
\eea
where we have used the anti-symmetry of the $f_{abc}$ in the first line, the structure relations to obtain the second line, and the definitions of the quadratic Casimir values  $C_2(G)$ and $C_2(r),$ respectively, in the adjoint representation and the representation $r$.

\sm

The Lie algebra factors for other diagrams can be computed in a similar manner. All the required integrals are simpler versions of the ones that appear in the gravitational case. We skip them here for brevity. Evaluating the Wilson line, we find the expected anomalous dimension to order $1/k^3$,
\bea
h={C_2(r)\over 2k}-{C_2(r)C_2(G)\over 4k^2}+{C_2(r)C_2(G)^2\over 8k^3}+\mathcal{O}(1/k^4)
\eea
which reproduces the expansion of the current algebra result in (\ref{pd}) to this order.


\section{Non-holomorphic  Wilson line from WZW}
\setcounter{equation}{0}
\label{sec:3}

In this section we discuss the computation of primary two-point functions in WZW models in a manner that does not exhibit manifest holomorphic factorization.  The virtue of this approach is that computations can be carried out using familiar dimensional regularization (modulo subtleties associated with epsilon tensors) with Feynman rules obtained from the WZW Lagrangian, and there is a simple relation between the bi-local primary operator and a Wilson line which holds even in the regulated theory.  The drawback is the lack of manifest holomorphic factorization, which in turn makes computations more laborious than those in the previous section, although the results are mutually consistent.

\subsection{Direct perturbative computation of $\langle g^{-1}(x_2) g(x_1)\rangle$ }

We proceed by computing $\langle g^{-1}(x_2) g(x_2)\rangle$  in perturbation theory, and then showing how this computation can be recast in terms of a non-holomorphic Wilson line. To carry out perturbation theory we parametrize the field $g(x)$ which takes values in the representation $r$ of the group $G$  in terms of the field $X(x)$ which takes values in the Lie algebra $\cG$ of $G$  \footnote{In this section repeated Lie algebra indices are summed over.}
\bea
\label{pf}
g(x) =\exp \left \{ {i\over \sqrt{k}}X^a (x)  T^a \right \}~.
\eea
Expanding the exponential in powers of $k^{-\half}$ and substituting into the WZW action yields
\begin{align}
\label{pg} S[g] &=  {1\over 8\pi} \int\! d^2x \, \p_\mu X^a \p^\mu X^a
+  {i\over 24\pi k^{1/2}} \, f_{abc}\int\! d^2x \eps^{\mu\nu}   X^a \p_\mu X^b \p_\nu X^c
\no \\
&~~~~-{1\over 24\pi k} \, K_{abcd}  \int\! d^2x \, X^a X^b \p^\mu X^c \p_\mu X^d  + \cO ( k^{-{3\over 2}}  )
\end{align}
where the metric is taken to be $ds^2 = dx^\mu dx^\mu$, and the tensor $K$ is given by
\bea\label{ph} K_{abcd}=  \Tr'\big(T^a T^b T^c T^d- T^a T^c T^bT^d\big)  = if_{bce}  \Tr'(T^a T^e T^d)~.
\eea
We work in dimensional regularization, taking the spacetime dimensionality to be $d=2-\ve$.  The one subtlety is how to define quantities involving $\eps^{\mu\nu}$ in this scheme; this will be discussed below. The position space free field correlator in $d=2 -\ep$ dimensions is given by
\bea
\label{pi}
\langle X^a(x) X^b(0)\rangle_{\rm free}  =  \Delta (x) \, \delta ^{ab}
\eea
where $\Delta(x)$ is the free-field propagator given by
\bea
\label{Delta}
\Delta (x) = \int {d^d p \over(2\pi)^d}  {4\pi \over p^2} e^{ipx}
= {\Gamma({d\over 2}-1) \over (\pi \, x^2) ^{{d\over 2}-1} }  =
 -{2\over \ve} - \ln(\pi \, x^2) - \gamma +\cO(\ve) ~.
\eea
and $x^2= \gamma_{\mu \nu} x^\mu x^\nu $ the the $d$-dimensional norm of $x^\mu$. As is familiar when using dimensional regularization, we are setting self-contractions to zero:  $\langle X^a(0)X^b(0)\rangle_{\rm free} =0$. Renormalizing the two-point function of the primary field $g$ to order $1/k$ by introducing a multiplicative renormalization factor $N(\ep) = 1 + 2 C_2(r) /( k \ep) + \cO(1/k^2)$, we find to this order
\bea \label{pj}
N(\ve) \langle g^{-1}(x)g(0)\rangle
=
N(\ve) \Big[  1 + {C_2(r)\over k}  \Delta (x)  + \cO (k^{-2} ) \Big]  =  1 - {C_2(r)\over k} \ln (x^2) + \cO(k^{-2})
\eea

\begin{figure}[h]
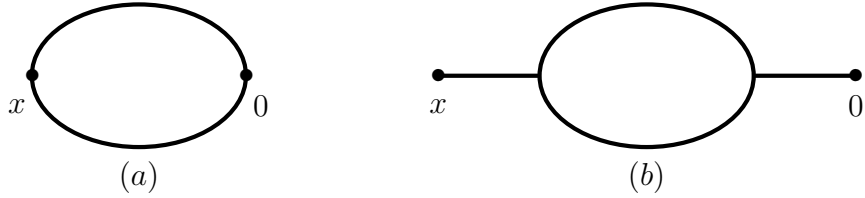

\begin{center}
\tikzpicture[scale=1.35]
\scope[xshift=-5cm,yshift=0cm]
\draw[ultra thick] (0,0)  ellipse (30pt and 20pt);
\draw[ultra thick] (5,0)  ellipse (30pt and 20pt);
\draw[ultra thick]  (6.06,0) -- (7.06, 0);
\draw[ultra thick]  (3.95,0) -- (2.95, 0);
\draw (7.06,0) node{$\bullet$};
\draw (2.95,0) node{$\bullet$};
\draw (-1.05,0) node{$\bullet$};
\draw (1.06,0) node{$\bullet$};
\draw (-1.2,-0.3) node{$x$};
\draw (1.2,-0.3) node{$0$};
\draw (2.95,-0.3) node{$x$};
\draw (7.06,-0.3) node{$0$};
\draw (0,-1) node{$(a)$};
\draw (5,-1) node{$(b)$};
\endscope
\endtikzpicture
\caption{Feynman diagrams at order $1/k^2$. \label{fig:1}}
\end{center}
\end{figure}

At order $1/k^2$ we have the diagrams shown in Figure \ref{fig:1}.   Figure \ref{fig:1}$a$ comes from expanding each of the exponentials in $g^{-1}(x)$ and $g(0)$ to second order and taking Wick contractions. Figure \ref{fig:1}b  arises from bringing down two cubic interaction vertices.   This yields
\bea
\label{pk}
\langle g^{-1}(x)g(0)\rangle_{1a} & = &  {1\over 2k^2} {\Gamma({d\over 2}-1)^2 \over \pi^{d-2}}  \Big( C_2(r)^2-{1\over 4} C_2(r)C_2(G)\Big) (x^2)^{2-d}~.
\no \\
\label{pl}
\langle g^{-1}(x)g(0)\rangle_{1b} & = &
-{1 \over k^2} {1\over 2^{d}\pi^{d-{5\over 2}} }{\Gamma({d\over 2}) \Gamma(d-2) \over (d-4)( d-2) \Gamma({d\over 2}+{1\over 2})} \, C_2(r)C_2(G)
\no \\ && \hskip 0.5in
\times
\gamma_{\mu\alpha} \eps^{\mu\nu} \eps^{\alpha\beta} [2(2-d)x_\nu x_\beta +x^2 \gamma_{\nu\beta}](x^2)^{1-d}~.
\eea
To proceed we need a rule for defining $\gamma_{\mu\alpha} \eps^{\mu\nu} \eps^{\alpha\beta}$ in d-dimensions.
 In $d=2$ we have
\bea
\label{pm}
\gamma_{\mu\alpha} \eps^{\mu\nu} \eps^{\alpha\beta} =\gamma^{\nu\beta}~.
\eea
One option is to adopt this rule in $d$ dimensions.  But there are alternative prescriptions as well.  For instance, we could first use the $d=2$ identity $ \eps^{\mu\nu} \eps^{\alpha\beta} =\gamma^{\mu\alpha}\gamma^{\nu\beta}-\gamma^{\mu\beta}\gamma^{\nu\alpha}$ and then contract with $\gamma_{\mu\alpha}$ in $d$ dimensions.  This gives $ \gamma_{\mu\alpha} \eps^{\mu\nu} \eps^{\alpha\beta}=(1-\ve)\gamma^{\nu\beta}$.  More generally, we could multiply $\ve$ by any coefficient.  These prescriptions differ in the sense that one can show that the value of the anomalous dimension depends on the value of this coefficient.  However, conformal invariance singles out the rule (\ref{pm}).  In particular, consider the current algebra Ward identity
\bea
\label{pn}
\langle g^{-1}(x_2)g(x_1) J^a(x_3)\rangle \sim \left({1\over z_3-z_2}- {1\over z_3-z_1} \right)\langle g^{-1}(x_2)g(x_1) \rangle T^a ~.
\eea
This Ward identity, together with the definition of the Sugawara stress tensor, is what fixes the conformal dimension of $g$ in the algebraic approach to the WZW model.  Evaluating both sides of (\ref{pn}) in $1/k$ perturbation theory we encounter, at order $1/k^2$, on the right hand side the same diagrams as above, including the ambiguity associated with the product of epsilon tensors.  On the other hand, no epsilon tensors appear on the left hand side at this order, and hence there is no ambiguity.  We then find that demanding (\ref{pn}) implies that we should adopt (\ref{pm}).  In fact, it turns out that under this rule (\ref{pn}) holds for all $d$.  This discussion of course raises the question as to the proper rule at higher loop orders, where higher powers of epsilon tensors will arise. There is a natural generalization of (\ref{pn}) in which one reduces all products of epsilon tensors directly in $d=2$, but whether this is compatible with the Ward identity at higher orders in $1/k$ is an open question that we do not address here.

\sm

Returning to (\ref{pl}) we now have
\bea\label{po} \langle g^{-1}(x)g(0)\rangle_{1b} =
 {C_2(r)C_2(G)\over k^2} { \Gamma ({d\over 2}-1 )^2 \over 8\pi^{d-2} (d-1)} \, (x^2)^{2-d}~,
\eea
where we used the Legendre duplication formula to simplify.

\sm

The computation of the renormalized correlator $N(\ve) \langle g^{-1}(x)g(0)\rangle$ also receives a contribution from the $1/k$ term in $N(\ve)$.  However, we need not consider this as it has no bearing on the computation of the anomalous dimension, since the latter can be extracted from $x\p_x \ln  \langle g^{-1}(x)g(0)\rangle$.

\sm

Collecting all contributions through order $1/k^2$ we have
\bea
\label{pp}
\langle g^{-1}(x)g(0)\rangle &=&  N(\ve) \Big[
 1 +{C_2(r)\over k} { \Gamma({d\over 2}-1) \over \pi^{d/2-1} } \, (x^2)^{1-{d\over 2}}  +{C_2(r)^2\over 2k^2} {\Gamma({d\over 2}-1)^2 \over \pi^{d-2}} \, (x^2)^{2-d}
 \no \\ && \hskip 0.5in
  - \, {C_2(r)C_2(G)\over k^2} {\Gamma({d\over 2}-1)^2 \over 8\pi^{d-2} } \, {d-2\over d-1} \, (x^2)^{2-d}\Big]
\eea
Since the scaling dimension $h$ is identified via $ \langle g^{-1}(x)g(0)\rangle \sim (x^2) ^{-2h}$ we can extract it as
\bea
\label{pq}
h = -{1\over 4}\,  \lim_{\ep \to 0} \, x\p_x \ln \<g^{-1}(x) g(0)\>
\eea
Plugging in  (\ref{pp}) we find
\bea
\label{pr}
h = {C_2(r)\over 2k} -{C_2(r)C_2(G)\over 4k^2} + \cO(k^{-3})
\eea
in agreement with the expansion of (\ref{pd}) to this order.

\subsection{Non-holomorphically factorized Wilson line}
\label{sec:nonhowil}

We can convert the bi-local primary operator considered above into a Wilson line type object by using the identity
\bea\label{ps} g^{-1}(x_2) g(x_1) = P \exp\Big\{-\int_{x_1}^{x_2} dy^\mu \, g^{-1}(y) \p_\mu g(y) \Big\}~.
\eea
This identity holds for any matrix-valued object $g(x)$.  In particular, if we compute the expectation value of both sides we are guaranteed to get exact agreement even with a finite regulator in place.  The computations of the previous section therefore establish that perturbation theory will yield $\big \langle P \exp\Big\{-\int_{x_1}^{x_2} g^{-1}(y) \p_\mu g(y) dy^\mu\Big\} \big \rangle \sim (x_{21}^2)^{-2h} $;  finiteness also requires the multiplicative renormalization factor $N(\ve)$ that we will suppress.

\sm

We now write
\bea
\label{pt}
\langle g^{-1}(x_2) g(x_1)\rangle =  \Big \<  P \exp\Big\{{1\over k} \int_{x_1}^{x_2} dy^\mu \, \Jc_\mu(y)  \Big\}  \Big \>
\eea
where the ``vector operator" $\Jc_\mu$ is defined as
\bea
\label{pu}
\Jc_\mu =-k g^{-1}\p_\mu g~.
\eea
This is not a conserved current, $\p^\mu \Jc_\mu \neq 0$.  Its components are related to those of the conserved currents $J_\mu$ and $\Jb_\mu$ as  $\Jc_z = J_z$, $\Jc_{\zb} = g^{-1} \Jb_{\zb} g$.   The  computations we have performed so far  establish that, as $\ve\rt 0$,
\bea
\label{pv}
\Big \<   P \exp\Big\{{1\over k} \int_{x_1}^{x_2} \!\! dy^\mu \, \Jc_\mu(y)  \Big\} \Big \>
=  \Big \<  P\exp\Big\{{\alpha\over k}\int_{z_1}^{z_2} \!\! dy \, J_z(y) \Big\} \Big \>
\Big \langle  P\exp\Big\{{\alpha\over k}\int_{\zb_1}^{\zb_2} \!\! d\overline{y}  \Jb_{\zb}(y))\Big\} \Big \rangle
\eea
through at least $\cO(1/k^2)$.  We note that the chiral Wilson lines on the right hand side require vertex renormalization factors, while no such object is required on the left hand side, as follows from the identity  (\ref{ps}).   Roughly speaking, we may surmise that the $\alpha$ factors on the right compensate for the non-chiral correlators on the left.

\sm

To flesh this out a bit more, let us consider  correlation functions involving the vector operator $\Jc_\mu$.  To order $k^0$ we find the two-point functions
\bea\label{pw}   \langle \Jc^a_z(x) \Jc^b_z(0)\rangle
& =&
{d\over 2}\left({d\over 2}-1\right)  {k\over z^2} \Delta(x)  \delta^{ab}~,
\cr
  \langle \Jc^a_z(x) \Jc^b_{\zb}(0)\rangle
  & =&
  \left({d\over 2}-1\right)^2 {k\over z\zb}\Delta(x) \delta^{ab}
  + { \left({d\over 2}-1\right)^2 \over 2(d-1)} C_2(G) {\Delta(x)^2 \over z\zb}  \delta^{ab}~,
  \cr
  \langle \Jc^a_{\zb} (x) \Jc^b_{\zb}(0)\rangle
  &= &
  {d\over 2}\left({d\over 2}-1\right)  {k\over \zb^2} \Delta(x)  \delta^{ab}
  +   \left({d\over 2}-1\right) C_2(G) {\Delta(x)^2 \over \zb^2 }   \delta^{ab}~,
  \eea
where $ \Delta(x)$ is the  scalar propagator defined in (\ref{Delta}).  The fact that $\langle \Jc_z \Jc_z\rangle $ is uncorrected at order $k^0$ is consistent with the fact that this is the two-point function of the conserved current $J_z$, and hence is unrenormalized.  The mixed correlator in the second line, $\langle \Jc_z \Jc_{\zb}\rangle$ is finite as $\ve\rt 0$, and this contributes to the non-chiral Wilson line expectation value at order $1/k^2$.  The correlator in the last line $\langle \Jc_{\zb}\Jc_{\zb}\rangle$ diverges as $\ve \rt 0$.    We define the renormalized operator $\Jct^a_{\zb}$,
\bea\label{px} \Jc^a_{\zb}= \left( 1-{C_2(G) \over k\ve} \right) \Jct^a_{\zb}~.
\eea
After doing this and taking $\ve \rt 0$ we  get
\bea\label{py}    \langle \Jc^a_z(x) \Jc^b_z(0)\rangle   & =& {k\over z^2}\delta_{ab}  +O(\ve)~, \cr
  \langle \Jc^a_z(x) \Jct^b_{\zb}(0)\rangle  & =&\cO(\ve)~,  \cr
  \langle \Jct^a_{\zb} (x) \Jct^b_{\zb}(0)\rangle   &=&A{k\over \zb^2}(z\zb)^{-{C_2(G)\over k}}\delta_{ab} +O(\ve)~, \eea
for some constant $A$.  $\Jct_{\zb}^a$ has acquired scaling dimension $(h,\overline{h}) = (1+{C_2(G)\over k},{C_2(G)\over k})$.   Coming back to the Wilson line, even if we rewrite it in terms of the renormalized vector operator components $(\Jc_z,\Jct_{\zb})$ it is not correct to omit the contributions from $\langle \Jc_z \Jct_{\zb}\rangle $ even though this correlator vanishes as $\ve \rt 0$. This vanishing is compensated by $1/\ve$ divergences, yielding a finite result.  Thus, there is  no manifest factorization.

\subsection{Comments on holomorphic factorization}

The conclusion of the above analysis is that the expectation value of the non-holomorphic Wilson line built out of $\Jc_\mu$ agrees with (the square of) the holomorphic Wilson line as the regulator is removed.   The former thus exhibits factorization to the order we have considered, but this comes out from detailed computation rather than being manifest from the start. Here we add a few more comments regarding this state of affairs.

\sm

The classical WZW model exhibits holomorphic factorization in the following sense.  The Euler-Lagrange equations are $\p_{\zb} (g^{-1} \p_z g)=0$.  The general solution of this equation takes the factorized form  $g(z,\zb) = \overline{g}_L(\zb) g_R(z)$, for arbitrary and independent (ignoring any reality conditions) matrices $\overline{g}_L(\zb)$ and $g_R(z)$.   Formally, the quantum correlator of interest is then
\bea
\label{zzy} \langle g^{-1}(z,\zb) g(0)\rangle = \langle g_R^{-1}(z)\overline{g}^{-1} _L(\zb)\overline{g}_L(0)g_R(0) \rangle~.\eea
We can try to argue for factorization from either the path integral or operator perspectives.  In terms of the path integral, we can imagine independently integrating over $g_L$ and $g_R$.  Inside  the $g_L$ path integral $\overline{g}^{-1} _L(\zb)\overline{g}_L(0)$ will be proportional to the unit matrix, and the correlator thence factorizes.  Of course, this argument is little more than handwaving as it stands, since the fact that classical solutions take the factorized form does not imply that we can perform independent path integrals over the two factors.  On the other hand, writing $g(z,\zb) = \overline{g}_L(\zb) g_R(z)$ makes more sense in the quantum theory if we work in the operator formalism.  In this case, the outstanding issue is that although the  oscillator modes can be uniquely associated to one of the two factors, the same is not true of the zero modes, which couples the two together.

\sm

We should also mention the argument by Witten \cite{Witten:1991mm} establishing the holomorphic factorization of current correlators on arbitrary Riemann surfaces, which is formal in the sense of ignoring UV divergences and anomalies.  Starting from the WZW action $S[g]$ one gauges the current $J$ by coupling to an external gauge field $A$,
\bea
\label{zzw}
S[g,A]=S[g]+{1\over 2\pi}\int d^2z \,  \Tr A_{\zb} g^{-1}\p_z g-{1\over 4\pi} \int d^2z \, \Tr A_{\zb}A_{z}~.\eea
The path integral over $g$ defines a wavefunction
\bea\label{zzx} \Psi(A) = \int\! Dg \, e^{-kS(g,A)} \eea
which serves as a generating function for current correlators.  The main result is to then show that the partition function, $Z(\Sigma)$, of the WZW model on the Riemann surface $\Sigma$ is equal to the norm of the wavefunction, $Z(\Sigma) = |\Psi|^2$,  where $|\Psi|^2 = {1\over {\rm Vol}(\hat{G})} \int\! DA \overline{\Psi(A)}\Psi(A)$.
We might contemplate extending this to our context by cutting holes in the Riemann surface with prescribed holonomies to represent the primary operator insertions.
Of course, one would still need to confront what for us is the main issue, namely making precise sense of these manipulations at the quantum level.  We leave these questions for the future, and now return to the main case of interest, the gravitational Wilson line.


\section{Renormalization of  gravitational Wilson lines}
\setcounter{equation}{0}
\label{sec:5}

In this section we shall regularize and renormalize  the matrix elements of the gravitational Wilson line operator in two-dimensional conformal field theory in a perturbative expansion in inverse powers of the central charge $c$. We focus on the scaling dimension $h (j,c)$ of the Wilson line operator, whose exact expression is predicted from the twisted $SL(2,\RR)$ current algebra representations of spin $j$ as discussed in the Introduction. Using the regularization and renormalization schemes developed here we shall calculate $h (j,c)$ up to order $1/c^3$ and find perfect agreement with the large $c$ expansion to the same order of the exact expression (\ref{aab}), which we repeat here
\bea
\label{hjc1}
h (j,c) = -j - { 6 \over c} j(j+1) -{78 \over c^2} j(j+1) -{1230 \over c^3} j(j+1) + \cO(c^{-4}).
\eea

As discussed in Section \ref{Wilsec} the Wilson line is defined as a matrix element of
\bea\label{zzz} P \exp \int _0 ^z d y \left ( L_1 + { 6 \over c} \, T(y) \, L_{-1} \right ).\eea
 The first step in implementing $1/c$ perturbation theory is to rewrite this in a manner analogous to what one does when passing to the interaction representation in quantum mechanical problems.  In the present case this amounts to using the identity
\bea
P \exp \int _0 ^z d y \left ( L_1 + { 6 \over c} \, T(y) \, L_{-1} \right )=
e^{z L_1} \, P \exp \int _0 ^z dy \left ( { 6 \over c} \, X(y) \, T(y) \right )
\eea
where $X(y)$ is given by,
\bea
\label{X}
X(y) = L_{-1} -2 y L_0 + y^2 L_1.
\eea
We shall consider matrix elements between states $|j,m\>$, with $2j+1 \in \NN$ and $0\leq j-m \leq 2j$, which are the tensor product of a  spin $j$ representation state of $SL(2,\RR)$ with  the ground state of the two-dimensional conformal field theory. In the infinite $c$ limit, the Wilson line operator reduces to $e^{z L_1}$ whose matrix element $\< j, -j| e^{z L_1} |j, j\> = z^{2j}$ gives the classical scaling dimension $-j$, in agreement with the leading term in  (\ref{hjc1}).

\sm

For large but finite $c$ we shall use perturbation theory in powers of $1/c$ to expand the Wilson line in terms of correlators which are polynomial in the stress tensor.  Such correlators may be evaluated on the two-dimensional plane  using the conformal Ward identities expressed, for example, in terms of the OPE of two stress tensors at points $w, z \in \CC$,
\bea
\label{TT}
T(z) T(w) = { c/2 \over (z-w)^4 } + { 2 \, T(w) \over (z-w)^2} +{ \p_w T(w) \over z-w} + \cO( (z-w)^0).
\eea
The perturbative expansion of the matrix elements of the Wilson line operator is beset by short distance singularities resulting from the first term in (\ref{TT}), and require regularization. The use of a Pauli-Villars regulator  in \cite{Besken:2017fsj}  correctly reproduced the $1/c$ term in (\ref{hjc1}) and the corresponding order $1/c^2$ term proportional to $(\ln z)^2$ in the expansion of the two point function, but gave a $1/c^2$ correction that disagrees with the corresponding term in (\ref{hjc1}).  Dimensional regularization, and analytic continuation in $\ep=2-d$ as applied to this problem  in \cite{Hikida:2017ehf}, is more successful, as we now discuss.

\subsection{Dimensional regularization}

No regulator of short distance singularities which preserves the infinite-dimensional conformal symmetry in two-dimensional space-time is known to exist. In fact most regulators will break the finite-dimensional conformal group and its dilation subgroup. However, dimensional regularization, in which the dimension of space-time is continued from two to $d=2-\ep$ dimensions, preserves dilation symmetry in  dimension $d$ in each Feynman diagram contribution for all values of $d$ where such diagrams are absolutely convergent. For this reason, dimensional regularization and analytic continuation in $\ep$ appears perhaps better-suited for regularizing correlators in scale invariant theories than other schemes.  Unfortunately,  the Ward identity (\ref{TT}), by which all correlators polynomial in the stress tensor can be computed on the two-dimensional plane, no longer holds and cannot be used to this end in $d \not=2$.

\sm

Therefore, we need a concrete {\sl quantum field theory representation} or {\sl model} for the stress tensor which is valid for arbitrary dimension $d$ and for arbitrary central charge $c$. Of course, upon proper renormalization, the Wilson line expectation values are expected to be independent of the model used to represent the CFT. To obtain an expansion for large $c$, we may take $c$ to be an integer, without loss of generality. A simple {\sl model} is then provided by the free field theory of $c$ scalar fields $\phi ^\gamma$ with $\gamma =1, \cdots, c$ in $d$ space-time dimensions.
Parametrizing space-time $\RR^d$ by coordinates $(z, \bar z,  \vec{z})$ where $z, \bar z$ are the complex coordinates for $\CC$ and $\vec{z} \in \RR^{d-2}$, we readily evaluate the normalized two-point function of the field $\p_z \phi^\gamma$,
\bea
\< \p_z \phi ^{\gamma}(z) \p_w \phi ^{\gamma '} (w)  \> = { - V(d) \,  \delta ^{\gamma \gamma '} (\bar z- \bar w)^2
\over \big ( |z-w|^2 +(\vec{z} - \vec{w})^2 \big )^{{d \over 2} +1}}~.
\eea
The normalization is given by $V(d) =\Gamma ({d\over 2} +1) /  \pi^{{d \over 2}-1}$, but we shall soon see that its effect may absorbed by a renormalization, and therefore we shall set $V=1$. For two points in the complex plane we have $\vec{z}=\vec{w}=0$, and for two points on the real line the correlator in $d=2-\ep$ dimensions  simplifies to the following formula we shall use throughout,
\bea
\< \p_w \phi ^{\gamma}(z) \p_z \phi ^{\gamma '} (w)  \> = { -   \delta ^{\gamma \gamma '}   \over  |z-w|^{2-\ep}}~.
\eea
In this model, the holomorphic stress tensor $T(z)$ for $z \in \CC$ is defined as the $T_{zz}$ component of the $d$-dimensional traceless stress tensor for the free field $\phi^\gamma$, which is given by,
\bea
T(z) = - \half \sum _{\gamma=1}^c : \p_z \phi ^\gamma (z) \p_z \phi^\gamma (z) :
\eea
where the normal ordering symbol $::$ instructs us to omit all self-contractions in the calculation of correlators of $T(z)$. An equivalent definition in terms of the OPE of two fields $\p_z\phi^\gamma$ may be given but will not be needed here.

\sm

Given the rules for calculating correlators in the free field theory model for the dimensionally regularized conformal field theory, it is straightforward to compute the correlator of the product of an arbitrary number of stress tensors, arranged at points  $y_i$  along the real line. Evidently, we have $\< T(y)\>=0$. The Feynman diagrams for a correlator $\< T(y_1) \cdots T(y_n)\>$ for $n\geq 2$ may be distinguished by  the number of connected  one-loop sub-diagrams.  Each sub-diagram  may be labelled by a {\sl partition $P$ into cycles} of the set of points $\{ y_1, \cdots, y_n\}$, with  each cycle  containing at least two points. Two partitions are equivalent if they are related by cyclic permutations and/or reversal of orientation of the points in each cycle, and under permutations of the cycles.  This partitioning of a Feynman diagram  into cycles is unique.

\sm

We shall denote a cycle of ordered points $y_{i_1} , \cdots , y_{i_\ell}$ by a square bracket $[i_1, \cdots, i_\ell]$
and the value of the corresponding one-loop diagram along this cycle by,
\bea
\label{T123}
\< T^2 \> _{[i_1, i_2]} & = & { c/2 \over |y_{i_1} - y_{i_2}|^{4-2\ep}}~,
\no \\
\< T^\ell \> _{[i_1, \cdots, i_\ell]} & = & { c \over |y_{i_1} - y_{i_2}|^{2-\ep} |y_{i_2} - y_{i_3}|^{2-\ep}
\cdots |y_{i_\ell} - y_{i_1}|^{2-\ep} }\, , \hskip 0.5in \ell \geq 3 ~.
\eea
The $y$-dependence of $\< T^\ell \> _{[i_1, \cdots, i_\ell]}$ is indicated through the indices $i_1, \cdots, i_\ell$ in the labeling of the cycle. The correlator  is given by a sum over all possible inequivalent partitions $P= C_1 \cup C_2 \cup \cdots \cup C_p$ into $p$ cycles, with $C_s \cap C_{s'} = \emptyset$ for $s' \not= s$, of the set $\{ y_1, \cdots, y_n\}$,
\bea
\label{TC}
\big \< T(y_1) \cdots T(y_n) \big \> = \sum _P \< T^n\> _P~,
\hskip 1in
\< T^n\> _P= \prod _{s=1}^p \< T^{\ell_s} \> _{C_s}~.
\eea
The $c$-dependence of the contribution of $P$ is given by $c^p$.
For the calculation of the matrix elements of the Wilson line operator to order $1/c^3$, to be given in the next section, we shall  need the correlators for $n=2,3$ given in (\ref{T123}),  as well as those for $n=4$ with one and two cycles, for $n=5$ with two cycles, and for $n=6$ with three cycles,  given as follows,
\bea
\label{tdec}
\<T(y_1) \cdots T(y_4)\> & = & \< T^4\> _{[12][34]}+ \< T^4\> _{[13][24]}+ \< T^4\> _{[14][23]}
\no \\ &&
+ \< T^4\> _{[1234]}+ \< T^4\> _{[1342]}+ \< T^4\> _{[1324]},
\no \\
\<T(y_1) \cdots T(y_5)\> & = &
\< T^5\> _{[12][345]} + \text{9 more partitions}  + \cO(c),
\no \\
\<T(y_1) \cdots T(y_6)\> & = &
\< T^6\> _{[12][34][56]} + \text{14 more partitions} + \cO(c^2).
\eea
The contributions from each partition is given by (\ref{TC}) and $ \< T^4\> _{[12][34]} =  \< T^2\> _{[12]} \< T^2\> _{[34]}$, $\< T^5\> _{[12][345]}= \< T^2\> _{[12]}\< T^3\> _{[345]}$, $\< T^6\> _{[12][34][56]}=\< T^2\> _{[12]}\< T^2\> _{[34]}\< T^2\> _{[56]}$ and their permutations.

\subsection{The regularized Wilson line matrix elements}

We define the regularized matrix element of the Wilson line operator in dimension $d=2-\ep$,
\bea
\label{W}
W _\ep(z) = N(\ep)  \< j, -j | e^{z L_1} \,
P \exp \left \{ { 6  \alpha (\ep)  \over c} \int _0 ^z dy \, X(y) \, T(y) \right \} |j,j\>.
\eea
 $X(y)$ was defined in (\ref{X}) and the states $|j,m\>$ stand for the tensor product of the free field theory ground state and the spin $j$ representation state of $SL(2,\RR)$ of weight $m$. The multiplicative renormalization  factor $N(\ep) $ is required on general grounds for an exponential operator, while the factor $\alpha (\ep) $ renormalizes the coupling to the stress tensor.

\sm

It will be shown below that the parameters $N(\ep) $ and $\alpha(\ep) $ may be chosen, order by order in powers of $1/c$, so as to cancel the poles in $\ep$, and to define a renormalized matrix element whose scaling dimension is $h(j,c)$,
\bea
\label{Wscaling}
W(z)  = \lim _{\ep \to 0} \, \<W_\ep(z)\> = z^{-2 h(j,c)}
\hskip 1in
z >0
\eea
up to order $1/c^3$ included. It will also be of interest to regularize and renormalize the matrix elements of the Wilson line operator multiplied by a single stress tensor $T(x)$ for $x\in \RR$,
\bea
\label{TW}
T_x W _\ep(z) =  N(\ep)  \< j, -j |  T(x) |x|^{4-2\ep} \, e^{z L_1} \,
P \exp \left \{ { 6  \alpha (\ep)   \over c} \int _0 ^z dy \, X(y) \, T(y) \right \} |j,j\>.
\eea
By inspecting the scaling behavior of the correlators involving $T(x)$, it is clear that the expectation value $\<T_xW_\ep(z)\>$ tends to a finite limit as $x \to \infty$ and defines a matrix element $\<T_\infty W_\ep(z)\>$ whose behavior is predicted from the dilation Ward identity,
\bea
\label{TWlim}
\lim _{\ep \to 0} \, \<T_\infty W_\ep (z)\> = h(j,c) \, z^2 \, W(z).
\eea
We verify that the parameters $N(\ep) $ and $\alpha (\ep) $ required to renormalize $W$ also renormalize $T_\infty W$, as may be expected on the basis of the dilation Ward identity in dimension $d=2-\ep$.

\subsection{Perturbative expansion}

$\<W_\ep(z)\>$ may be evaluated by expanding the path ordered exponential in powers of $\alpha /c$,
\bea
\label{Wep}
\<W_\ep (z)\> = z^{2j} N  \sum _{n=0}^\infty  { (6 \alpha  )^n  \over c^n}
\int_{0}^z \!\! dy_n   \cdots \int_{0}^{y_2} \! \! dy_{1}
F_n (z;y_n, \cdots, y_1) \big \< T(y_n)  \cdots T(y_1) \big \>
\quad
\eea
where we have suppressed the $\ep$-dependence of $N$ and $\alpha$, which will be understood throughout. The $SL(2,\RR)$ group theory factor $F_n$ is defined by,
\bea
z^{2j} F_n (z;y_n, \cdots, y_1) = \< j,-j| e^{zL_1} X(y_n) \cdots  X(y_1) | j, j\>.
\eea
A recursive formula for $F_n$ is obtained in Appendix \ref{sec:A}, while the calculations of the stress tensor correlators were given in the preceding section. To proceed further, it will be convenient to organize the calculation of $\<W_\ep(z)\>$ as follows,
\bea
\label{Wex}
\<W_\ep (z)\>= z^{2j} N \sum _{n=0}^\infty  \alpha^n  \, z^{n \ep} \, W_{1\cdots n}
\eea
where $W_0=1$, $W_1=0$ and the contributions for $n \geq 2$ are given by,
\bea
\label{Wn}
W_{1\cdots n} =  {6^n \over  c^n \, z^{n \ep}} \int _0 ^z dy_n  \cdots \int ^{y_2}_0 dy_1
F_n(z;y_n, \cdots, y_1) \big \< T(y_n)  \cdots T(y_1) \big \>.
\eea
The factors of $z^{n \ep}$ have been inserted  to make the coefficients $W_{1\cdots n}$ independent of $z$ for any value of $\ep$.  To see this, we recall from Appendix \ref{sec:A} that the combination $z^nF_n (z;y_n, \cdots, y_1)$ is a homogeneous polynomial  in $z, y_1 , \cdots, y_n$ of total degree $2n$, while the correlator of $n$ stress tensors is homogeneous in $y_1, \cdots, y_n$ of total degree $n(-2+\ep)$. Therefore $W_{1 \cdots n}$ is homogeneous in $z, y_1 , \cdots, y_n$ of total degree 0 and we may set $z=1$ in the evaluation of $W_{1 \cdots n}$ in (\ref{Wn}) so that all $z$-dependence of $W_\ep(z)$ resides in the coefficients $z^{n \ep}$ in (\ref{Wep}).
The expansion of $\<T_\infty W_\ep (z)\>$ proceeds analogously by replacing the correlator $\big \< T(y_n)  \cdots T(y_1) \big \>$ with $\big \< T(x) |x|^{4-2\ep} T(y_n)  \cdots T(y_1) \big \>$ and then taking the $x \to \infty$ limit.

\sm

The coefficients $W_{1 \cdots n}$ may be decomposed into a sum over inequivalent partitions $P$
 of the set of $n$ points $\{ y_1, \cdots, y_n\}$ by decomposing the correlator of $n$ stress tensors in (\ref{Wn})
 into a sum over $P$ using (\ref{TC}),
\bea
W_{1\cdots n} = \sum _P W_P,
\hskip 0.6in
W_P  =  {6^n \over c^n} \int _0 ^1 dy_n \cdots \int ^{y_2}_0 dy_1
F_n(1;y_n, \cdots, y_1) \< T^n \>_P.
\eea
The  expression for $W_P$ may be simplified using  the scaling and translation properties of $\< T^n \>_P$ and the polynomial nature of the function $F_n (1;y_n, \cdots, y_1)$ to resolve the nested ordering of the integrals. We change variables from $(y_n, \cdots, y_1)$ to $(x_n, u, \a_{n-1}, \cdots, \a_1)$,
\bea
\label{yau}
y_k & = &  x_n - u \alpha_{n-1} - u \alpha _{n-2} \cdots - u \alpha _k  \hskip 1in   1 \leq k \leq n-1,
\no \\
y_n & =  & x_n
\eea
subject to $0 \leq u \leq x_n \leq 1$ and $0 \leq \alpha _i$ as well as $\alpha_{n-1} + \cdots \alpha _1=1$.
Using the observation that the integration range of the variables $u,x_n$ is independent of the integration range of the variables $\alpha_i$, we rearrange the integrations as follows,
\bea
\label{wpar}
W_P = { 6^n \over 2^{p_2} c^{n-p}}   \int _0 ^1 d\alpha_{n-1} \cdots \int ^1_0 d\alpha _1 \,
\delta \left (1-\sum_{k=1}^{n-1} \alpha_k \right )
{ \cN_n (\alpha_1, \cdots, \alpha _{n-1}) \over \cD_P (\alpha_1, \cdots, \alpha _{n-1})}~.
\eea
The function $\cD_P$ is given in terms of the contribution to the stress tensor correlator arising from the partition $P$ and is given explicitly by,
\bea
\label{tnp}
\< T^n \>_P = { c^p \over 2^{p_2}} { u^{-2n + n \ep}  \over \cD _P (\alpha _1, \cdots, \alpha _{n-1})}
\eea
where $p$ is the total number of cycles in $P$ and $p_2$ is the number of 2-cycles in $P$.
The function $\cN_n$ is defined as follows,
\bea
\label{nn}
\cN_n (\alpha _1, \cdots , \alpha _{n-1}) =  \int _0 ^1 du \, u^{-n-2+ n \ep} \int _u ^1 dx_n \,
F_n(1;y_n, \cdots, y_1)
\eea
where $y_1, \cdots y_n$ are given in terms of $x_n, u, \alpha _1, \cdots \alpha _{n-1}$ by  (\ref{yau}). Since $F_n(1;y_n, \cdots, y_1)$ is polynomial in $y_i$, the integral $\cN_n$ is polynomial in $\alpha _i$ as well, with coefficients which are rational functions of $\ep$.  Finally, one of the $\alpha _k$-integrals in (\ref{wpar}) may be carried out by satisfying the $\delta$-function, so that the number of non-trivial integrals left over is $n-2$.

\subsection{Evaluation of $W_{1 \cdots n}$}

The details of the calculation of the functions $W_P$ and their sum $W_{1\cdots n}$ are presented in Appendix \ref{sec:B}. They include the list of the denominator functions $\cD_P$ and the evaluations of some of the integrals over the parameters $\alpha _i$, but we do not give the functions $F_n$ or $\cN_n$ whose length grows rapidly with $n$ and were handled by MAPLE. The result may be summarized as follows. The contribution $W_{12}$ is of order $1/c$ and is required up to order $\ep^2$, the contribution  $W_{123}$ is of order $1/c^2$ while $W_{1234}= W_{1234}^{(2)}+ W_{1234}^{(3)}$ has contributions of order $1/c^2$ and $1/c^3$ and both are required to order $\ep^0$,
\bea
\label{Wup4}
c \,  W_{12}  & = & { 6j(j+1) \over \ep} + j(10j+4)+{j \over 3} (74j+98)\ep +{ j \over 9} (418j + 196) \ep^2,
\no \\
c^2 \, W_{123} &= &
-{ 96j(j+1) \over \ep^2} + {24 j \over \ep} (2j^2-9j-5)
 + 16\pi^2 j(j+1)    + 6j(18j^2-143 j -203),
\no \\
c^2 \, W_{1234}^{(2)} &= &
{18  \over \ep^2}j (j+1) (j^2 + j + 2) + {3 \over \ep}j (20 j^3 + 16 j^2 + 49 j + 29 )
\no \\ &&
\hskip 0.3in
+2 j (99 j^3 + 132 j^2 + 436 j  + 460)     -24 j (j+1) \pi^2,
\no \\
c^3 \, W_{1234}^{(3)} &= &
{1296 \over \ep^3}j(j+1) + {648  \over \ep^2}j(- 2 j^2 + 5 j + 3)
+ {216   \over \ep}j (2 j^3 - 11 j^2 + 89 j  + 132 )
\no \\ &&
\hskip 0.3in
+{72\over5} j (4 j^3 + 8 j^2  - 39 j   -43 ) \pi^2.
\eea
Finally, the contributions $W_{12345}$ and $W_{123456}$ are required to order $1/\ep$ and to order $1/c^3$, for the calculation of the dimension $h(j,c)$ to order $1/c^3$,
\bea
\label{Wup6}
c^3 \, W_{12345} &=&
-{576\over 5\ep^3} j (j+1) (5 j^2 + 5 j+11)  + {96j\over 5\ep} j (1 + j) (5 j^2 + 5 j+79) \pi^2
\no \\ && \hskip 0.3in
+ {48j\over 5\ep^2}  (30 j^4 - 205 j^3 - 152 j^2 - 634 j  -387  )
\no \\ && \hskip 0.3in
+ {4j\over 5\ep}  (1410 j^4 - 12341 j^3 - 18640 j^2 - 58776 j  -62077  ),
\no \\
c^3 \, W_{123456} &=&
{36 j\over \ep^3} (j+1) (j^2 + j + 2) (j^2 + j + 4) -{48 j\over \ep} (j+1) (3j^2+3j+13) \pi^2
\no \\ && \hskip 0.3in
+ {6 j\over \ep^2} (30 j^5 + 36 j^4 + 201 j^3 + 210 j^2 + 361 j  + 202  )
\no \\ && \hskip 0.3in
+ {2 j\over \ep} (372 j^5 + 468 j^4 + 3873 j^3 + 5967 j^2 + 10100 j  + 8684  ).
\eea
The calculation of $\<T_\infty W_\ep(z)\>$ is analogous. The results are given in the Appendix~\ref{sec:D3}.

\subsection{Renormalization of $W_\ep(z)$ and $T_\infty W_\ep(z)$ to order $1/c^3$}

To order $1/c^3$,  the regularized matrix element $\<W_\ep(z)\>$ of the Wilson line operator is given by (\ref{Wex}), (\ref{Wup4}), and (\ref{Wup6}), as well as by the parameters $N$ and $\alpha $. We seek to determine $N$ and $\alpha $ by requiring that $\<W_\ep (z)\>$ obey as renormalization conditions the scaling relation (\ref{Wscaling}) to order $1/c^3$. By inspecting the expansion of $\<W_\ep(z)\>$ in terms of the coefficients $W_{1 \cdots n}$ it is far from obvious that such a scaling relation can indeed be secured. However, once it has been, the parameter $N$ is trivially  fixed as follows,
\bea
\<W_\ep(1)\>=1.
\eea
This leaves the parameter $\alpha$ at our disposal to enforce the scaling relation (\ref{Wscaling}) by requiring that the function $\ln \<W_\ep(z)\>$ be linear in $\ln (z)$,
\bea
\ln \<W_\ep (z)\> = - 2 h(j,c) \ln z + \cO(\ep)
\eea
where $h(j,c)$ is to be determined in the process. By inspecting the relation between the order of expansion in powers of $1/c$ and the order of the pole in $\ep$, we find that for order $1/c^m$ the maximal order is $1/\ep^m$, thereby producing a polynomial in $\ln(z)$ of degree $m$ in $\ln\< W_\ep(z)\>$, up to corrections of order $\cO(\ep)$. Therefore, to order $1/c$, the scaling condition is automatic, while to orders $1/c^2$ and $1/c^3$ the scaling condition imposes respectively two and three conditions. These conditions are satisfied by a function $\alpha $ given as follows,
\bea
\alpha  = 1 + {1 \over c} \left ( {6 \over \ep} + 3 + \ep a _1  \right )
+{1 \over c^2} \left ( { 30 \over \ep^2} + {55 \over \ep} + a _2 + \ep a _3  \right ) +\cO(c^{-3}, \ep^2).
\eea
The contributions proportional to $a_1, a _2$ and $a_3$ are not determined by the renormalization scaling conditions, and neither are higher order terms in $1/c$ or $\ep$ to this order in the expansion.  The scaling dimension resulting from the renormalization of $W$ is given by,
\bea
h_W(j,c) = -j  - j(j+1) \left ( {6 \over c} + { 78 \over c^2} +
{ 60 a_2 - 360 a_1 + 2450  + 192 \pi^2 (3j^2+3j-1) \over 5 c^3} \right )
\eea
up to contributions of order $1/c^4$ and $\ep$. While the result for $h_W(j,c)$ to the orders $1/c$ and $1/c^2$ are uniquely determined by the renormalization procedure and precisely agree with the predictions of $SL(2,\RR)$ current algebra in (\ref{hjc1}), the order $1/c^3$ is determined only once the particular combination $a _2-6 a _1$  of the coefficients $a_1$ and $a _2$ is known.

\sm

The missing information may be obtained from the renormalization of the matrix element $T_\infty W_\ep(z)$. Its detailed calculation is given in the Appendix. Using the same renormalization parameters $N$ and $\alpha $ as we used for $W_\ep(z)$, the prediction of the scaling dimension derived from $\<T_\infty W_\ep(z)\>$ is obtained via (\ref{TWlim}) and is given by,
\bea
h_{TW}(j,c) = -j  - {6 j(j+1) \over c} - {j \over  c^2} \left (
 78 j + {49 \over 3}  + {16 \over 5} \pi^2   \big (3j(j+1) -1 \big ) -6a _1 +a_2   \right ).
\eea
Matching the orders in $1/c^2$ gives the following result for the combination,
\bea
\label{a2a1}
a_2-6 a _1= {185 \over 3}  - { 1 6 \pi^2 \over 5} \big ( 3 j(j+1) -1 \big )
\eea
which upon substitution in the $1/c^3$ term of $h_W(j,c)$ leads to perfect agreement with the predictions of (\ref{hjc1}) to order $1/c^3$.

\sm

We note that renormalization of the gravitational Wilson line matrix elements consistent with the conformal Ward identities has forced us to make the vertex renormalization parameter $\alpha (\ep)$ dependent on $j$ in the  order $1/c^3$ contribution to the Wilson line, and to order $1/c^2$ in $\alpha (\ep)$. This $j$-dependence of $\alpha(\ep)$ is a new phenomenon that was absent at lower orders in $1/c$, and raises two issues. First, in terms of renormalization theory, it suggests that the gravitational Wilson line operator as originally defined cannot be renormalized at the  operator level, since a dependence on the states governing its matrix elements enters. A slight modification of the original definition of the Wilson line can remedy this obstacle by promoting $\alpha (\ep) $ itself to an operator which involves the quadratic Casimir of $SL(2,\RR)$. Second, to satisfy (\ref{a2a1}), we actually have a choice:  setting $a_1=0$ we require a $j$-dependent renormalization at order $1/c^2$, while setting $a_2=0$ we can get away with a renormalization at order $1/c$ of an {\sl evanescent operator} which, given its proportionality to $\ep$, would vanish at the classical level as $\ep \to 0$. The role of such evanescent operators remains to be understood.


\section{Regularization scheme in two dimensions}
\setcounter{equation}{0}
\label{sec:6}

Instead of ``changing the theory" by extending the free field model for a conformal field theory with central charge $c$ from two dimensions to $d=2-\ep$ dimensions, we shall attempt in this section to keep conformal invariance intact in $d=2$, and regularize and renormalize the operator $W$ in this exactly conformal theory. As we shall show below, for the particular though natural regulator we choose, this attempt will ultimately fail.

\subsection{A two-dimensional  regulator for the Wilson line}

We introduce a regulator, order by order in the $1/c$ expansion of the matrix elements of the Wilson line operator, in which  the correlator of stress tensors $\< T^n \> _{1\cdots n}$  is regularized by,
\bea
\label{TnG}
\< T^n \> _{1\cdots n} = \< T(y_1) \cdots T(y_n) \> \prod _{1 \leq i < j \leq n} |y_j - y_i|^\ep
\eea
and the correlator $\< T(y_1) \cdots T(y_n) \>$ is evaluated using the OPE for the stress tensor of (\ref{TT}) of a conformal field theory with central charge $c$, valid strictly in two dimensions.  We have chosen the regulator to be symmetric under permutations of the points $y_1, \cdots, y_n$ just as the stress tensor correlator is, to be invariant under translations of the variables $y_i$, and to have good scaling behavior similar to, but different from, dimensional regularization.   In the $\tilde \alpha /c$ expansion, and with the regularization defined above, the Wilson line correlator may be presented as a sum over contributions with a definite number of $T$-insertions,\footnote{Throughout this section, we shall use a tilde for the quantities defined with the regulator of (\ref{TnG}) in order to distinguish them from those defined in the preceding section with dimensional regularization. }
\bea
\<\W_\ep (z)\>= z^{2j} \tilde N \sum _{n=0}^\infty \tilde \alpha ^n \, z^{\half n (n-1) \ep} \, \W_{1\cdots n}~.
\eea
The factors of $z^{\half n (n-1) \ep}$ have been extracted in order to make the coefficients $\W_{1\cdots n}$ independent of $z$, using arguments analogous to the ones used for $W_\ep(z)$.
The decomposition of the correlator into a sum over contributions arising from inequivalent partition cycles $P$ proceeds as with dimensional regularization, and we  have,
\bea
\W_{1\cdots n} = \sum _P \W_P,
\hskip 0.6in
\W_P  =  { 6^n \over c^n} \int _0 ^1 dy_n  \cdots \int ^{y_2}_0 dy_1 F_n(1;y_n, \cdots, y_1) \< T^n \>_P
\eea
where $\< T^n \>_P $ is defined by (\ref{TnG}) for the partition $P$.

\sm

Using the change of variables (\ref{yau}) we recast the expression for $\W_P$ as follows,
\bea
\W_P = { 6^n \over 2^{p_2} c^{n-p}}   \int _0 ^1 d\alpha_{n-1} \cdots \int ^1_0 d\alpha _1 \,
\delta \left (1-\sum_{k=1}^{n-1} \alpha_k \right )
{ \tilde \cN_n (\alpha_1, \cdots, \alpha _{n-1}) \over \tilde \cD_P (\alpha_1, \cdots, \alpha _{n-1})}
\eea
where $\tilde \cD_P$ and $\tilde \cN_P$ are defined by,
\bea
\< T^n \>_P  & = &  { c^p \over 2^{p_2}}
{ u^{-2n + \half n (n-1) \ep}  \over \tilde \cD _P (\alpha _1, \cdots, \alpha _{n-1})}~,
\no \\
\tilde \cN_n (\alpha _1, \cdots , \alpha _{n-1}) &  = &  \int _0 ^1 du \, u^{-n-2+ \half n (n-1) \ep} \int _u ^1 dx_n \,
   F_n(1;y_n, y_{n-1}, \cdots, y_1)
\eea
with $\< T^n \>_P$ given in (\ref{TnG}), $p$ and $p_2$ are respectively the total number of cycles and the number of two-cycles in $P$.

\subsection{Calculation of the coefficients $\W_{12}$, $\W_{123}$ and $\W_{1234}$}

The coefficient $\W_{12}$ coincides with the coefficient $W_{12}$ computed in dimensional regularization after letting $2\ep \to \ep$, while $\W_{123}=W_{123}$, and are given by,
\bea
c \,  W_{12} & = & { 12j(j+1) \over \ep} + j(10j+4)+{j \over 6} (74j+98)\ep  + \cO(\ep^2),
\no \\
c^2 \, W_{123}  & = &  -{ 96j(j+1) \over \ep^2} + {24 j \over \ep} (2j^2-9j-5)  + \cO(\ep^0).
\eea
To order $1/c^2$, the coefficient  $\W_{1234}$ receives contributions from the partitions $[12][34]$, $[13][24]$ and $[14][23]$, whose denominator functions  are given by,
\bea
\tilde \cD_{[12][34]} & =  &
 \a_1^{4-\ep} \, \a_2^{-\ep} \, \a_3^{4-\ep} \, (\a_1+\a_2)^{-\ep} \, (\a_2+\a_3)^{-\ep},
 \no \\
\tilde \cD_{[13][24]} & =  &
 \a_1^{-\ep} \, \a_2^{-\ep} \, \a_3^{-\ep} \, (\a_1+\a_2)^{4-\ep} \,  (\a_2+\a_3)^{4-\ep} ,
 \no \\
\tilde \cD_{[14][23]} & =  &
 \a_1^{-\ep} \, \a_2^{4-\ep} \, \a_3^{-\ep} \, (\a_1+\a_2)^{-\ep} \, (\a_2+\a_3)^{-\ep}.
\eea
The function $\tilde \cN_4(\alpha_1, \alpha_2, \alpha_3) $ is a polynomial in its variables, with coefficients which are rational functions of $\ep$ with simple poles. We satisfy the $\delta$-function constraint by solving for $\alpha_2=1-\alpha_1-\alpha_3$, and decompose the polynomial $\tilde \cN_4$ in the following, equivalent ways,
\bea
\tilde \cN_4 (\alpha_1, 1-\alpha_1-\alpha_3, \alpha _3)
& = & \sum_{A,B=0}^2  \cM^{(1)} _{AB} \, \alpha _1 ^A \alpha _3^B
= \sum_{A,B=0}^2  \cM^{(2)} _{AB} \, (1-\alpha_1)  ^A (1-\alpha _3)^B
\no \\
& = & \sum_{A=0}^4 \sum_{B=0}^2 \cM^{(3)} _{AB} \, (1-\alpha _1) ^A (1-\a_1-\a_3)^B.
\eea
The expansion reduces the integrals to sums over basic families of integrals $\cQ^{(i)}_\ep $ for $i=1,2,3$ given and evaluated in Appendix \ref{sec:D4},
\bea
\label{c2}
c^2 \, \W_{[12][34]}  & = & { 6^4  \over 4 }   \sum_{A,B=0}^2
\cM^{(1)} _{AB} \,\cQ^{(1)} _\ep (A-3,B-3),
\no \\
c^2 \, \W_{[13][24]}  & = & { 6^4  \over 4 }   \sum_{A,B=0}^2
\cM^{(2)} _{AB} \,\cQ^{(2)} _\ep (A-3,B-3),
\no \\
c^2 \, \W_{[14][23]}  & = & { 6^4  \over 4 } \sum_{A=0}^4 \sum_{B=0}^2
\cM^{(3)} _{AB} \, \cQ^{(3)} _\ep (A+1,B-3).
\eea
The results are as follows,
\bea
\label{c2}
c^2 \, \W_{[12][34]}  & = & { 56 \over \ep^2} j^2(j+1)^2 +{ 2 \over 15 \ep} j(j+1)(776j^2-1924 j +273),
\no \\
c^2 \, \W_{[13][24]}  & = & - { 16 \over \ep^2} j(j+1)(j^2+j-1) -{ 2 \over 15 \ep} j(466 j^3 + 1292 j^2 -21 j - 487),
\no \\
c^2 \, \W_{[14][23]}  & = & { 8 \over \ep^2} j^2(j+1)^2 +{ 4 \over 3 \ep} j(j+1)( 29 j^2 + 119 j - 69)
\eea
giving a combined contribution of
\bea
c^2 \, \W_{1234} = { 16 \over \ep^2}j(j+1)(3 j^2 + 3 j +1) + { 4 j \over 3 \ep} (60 j^3 -96 j^2 -113 j + 7).
\eea
Expanding the parameter $\tilde \alpha$ in powers on $1/c$,
\bea
\tilde \alpha (\ep) = 1 + { 1 \over c} \left ( {\tilde a _1 \over \ep} + \tilde a _2 \right ) + \cO(c^{-2}),
\eea
setting $\<\tilde W _\ep (1)\>=1$ and collecting all remaining contributions, we find,
\bea
\ln \<\W _\ep (z)\> & = & 2 j \ln z + {12 \over c}  j(j+1) \ln z
+ {24 \over c^2 \, \ep} j(j+1) (6 j^2 + 6 j -8 + \tilde a _1) \ln z
\no \\ &&
+{ 4j \over c^2} (60 j^3 -240 j^2 - 412 j -76  +5 j \tilde a _1 + 6 j \tilde a _2 + 2 \tilde a _1 + 6 \tilde a _2) \ln z
\no \\ &&
+ { 12 \over c^2 } j(j+1)(60 j^2 + 60 j -12 + \tilde a _1) (\ln z)^2 + O(\ep).
\eea
To obtain a finite result, we must cancel the pole in $\ep$ and thus set $\tilde a _1 = 8 - 6 j(j+1)$. Having done so, the value of the coefficient of $(\ln z)^2$ becomes $24 j(j+1)(27 j^2+27 j -2)$ and no further adjustment of $\tilde N$ or $\tilde \alpha$ is available to cancel this obstruction to the scaling behavior of (\ref{Wscaling}) for $\<\W_\ep(z)\>$.


\section{Discussion and Outlook}
\setcounter{equation}{0}
\label{sec:7}

 The main result of the paper is the computation of the expectation value of the gravitational Wilson line to order $1/c^3$. To deal with the short-distance singularities which arise in the integrations over  stress tensor correlators, we have used a version of dimensional regularization to dimension $d=2-\ep$ combined with a non-trivial analytic continuation in $\ep$, and effectively treated the stress tensor as having dimension $d=2-\ve$.
Renormalization of the gravitational Wilson line matrix elements consistent with the conformal Ward identities was found to require, to order $1/c^3$ included, an overall multiplicative factor $N(\ep)$ and a ``vertex renormalization" factor $\alpha(\ep)$. The multiplicative factor $N(\ep)$ depends on $\ep$ and $j$ in an expansion in powers of $1/c$. The vertex renormalization $\alpha(\ep)$ is independent of $j$ to orders $1/c$ and $1/c^2$ but requires dependence on $j$ through its Casimir value $j(j+1)$ to order $1/c^3$. This result suggests that, to sufficiently high order in $1/c$, the renormalization of the Wilson line operator depends on the matrix element considered. Deepening the understanding of this dependence is left for future work.

\sm

From a purely diagrammatic point of view, the emergence of a bi-local conformal primary operator from the gravitational Wilson line matrix elements appears to be based on the magic of remarkable relations between contribution at different orders in $1/c$.  For example, a simple fact about the anomalous dimension \eqref{aab} is that it depends on $j$ only through  the $SL(2,\mathbb{R})$ Casimir eigenvalue $j(j+1)$. Yet each diagram by itself does produce higher powers of $j$ which do not form a polynomial in $j(j+1)$.  No regularization scheme appears to be known in which each contribution is polynomial in $j(j+1)$.

\sm

As a simpler example, we have computed the expectation values of Wilson line operators of holomorphic currents  appearing in theories with level $k$ current algebra symmetry to order $1/k^3$. The computations are relatively simpler in this case but still retain a lot of the features of the gravitational case. We have also performed a more standard field theoretic perturbative calculation of the expectation value of a Wilson line for non-holomorphic currents using the WZW model to order $1/k^2$.  The results of the two approaches are consistent;  however the connection between the two calculations remains to be fully elucidated.

\sm

A promising approach towards a more geometrical understanding of the bi-local and conformal primary nature of gravitational Wilson lines is via  Hamiltonian reduction, which produces  Virasoro symmetry from $SL(2,\RR)$ current algebra symmetry (see \cite{Bershadsky:1989mf} for details). The constraints we need to impose on the $SL(2,\RR)$ currents $J^a(z)$ are given by $J^-(z)=k$ and $J^0(z)=0$. Under these constraints, the current algebra Wilson line reduces to the gravitational Wilson line (with central charge $c=6k$)
\bea
J^aT^a\longrightarrow L_1+{6\over c}L_{-1}~.
\eea
Further, it was shown in \cite{Alekseev:1988ce}, that the geometric action can be obtained from the chiral WZW action by the same reduction. The geometric action is written in terms of the function $f(z)$ appearing in \eqref{ai} and \eqref{aj}, and is the right object to compute stress tensor correlators. Note that the same reduction is done in the bulk Chern-Simons theory when we impose asymptotically $AdS$ boundary conditions. As a consequence, at least formally, the expectation value of the gravitational Wilson line can be obtained by reduction of the $SL(2,\RR)$ current algebra Wilson line
\bea
\label{disc1}
\int\mathcal{D}g \, e^{-\text{S}_{\text{WZW}}[g]} \, W(z) \longrightarrow
\int  \cD  f \, e^{-\text{S}_{\text{G}}[f]} \, W(z)~.
\eea
All this suggests that understanding the current algebra Wilson line might be sufficient to understand the gravitational case. However, the transformation from $g$ to $f$ in (\ref{disc1}) remains formal. Addressing the subtle issues of regularization and renormalization of the transformation, and the emergence of conformal symmetry, are left for future work as well.

\sm

Recently, the connection between the geometric action and $AdS_3$ gravity was carefully studied in \cite{Cotler:2018zff}. The authors used the geometric action and certain bi-local operators to calculate various quantities, such as the sphere and torus partition functions and corrections to Virasoro blocks. The bi-local operators used in \cite{Cotler:2018zff} are simply the Wilson line operators we consider (compare equation 6.9 there with \eqref{aj} here). It would be interesting to see if their methods could be used to understand our problem better.

\sm

The advantage of our regulator over, for example, the Pauli-Villars type regulator used in \cite{Besken:2017fsj} is that it is dimensionless. This greatly constrains the form of the divergences and allows a simple prescription to subtract divergences. Another natural dimensionless regulator was considered in section \ref{sec:6}. Surprisingly, we found that it is not possible to restore conformal invariance in this case, as we take the regulator away. Understanding why dimensional regularization is superior might shed some light onto the renormalization problem.

\sm

By computing the Wilson line anchored on the boundary, we are computing the boundary to boundary scalar two point function in $AdS_3$ with graviton loop corrections (up to 3 loops). A conventional calculation would be quite complicated as we would have to use the bulk to bulk graviton propagator and involves integrating vertices over all of $AdS$. The Wilson line calculation is manifestly holomorphically factorized and needs only one integration per vertex. This is much simpler. It would be interesting to see if we could reduce the standard Witten diagram computation to the Wilson line one.

\sm

Ultimately, we are interested in finding a formalism that allows us to exploit Virasoro symmetry to understand non-perturbative gravity corrections in $AdS_3$. We believe that understanding the renormalized Wilson line better is a step towards this direction.

\subsection*{Acknowledgments}

ED is grateful to Constantin Bachas for useful conversations and would like to thank DAMTP in Cambridge, LPTHE at Jussieu, and LPTENS at the Ecole Normale Sup\'erieure in Paris for their hospitality while part of this work was being completed. MB and AH are happy to acknowledge support from the Mani L. Bhaumik Institute for Theoretical Physics. The research of ED and PK is supported in part by the National Science Foundation under research grant PHY-16-19926 and ED was also supported in part by  a Fellowship from the Simons Foundation.

\appendix


\section{Alternative approach to $SL(2,\RR)$ matrix elements}
\setcounter{equation}{0}
\label{sec:AA}

In the bulk of this paper we based the Wilson line on finite dimensional spin $j$ representations of $SL(2,\RR)$.   These representations are convenient to work with, but since they are non-unitary one must analytically continue in $j$ at the end of any computation to obtain results valid for unitary representations.  Here we discuss an alternative approach that works with unitary representations throughout.

To this end we use the isomorphism between $SL(2,\RR)$ and $SU(1,1)$ to realize their unitary highest weight representations on the space of square-integrable  holomorphic functions on  the unit disk $D= \{ u \in \CC, |u|<1 \}$ with the  inner product \cite{Bargmann:1946me},
\bea
\label{xbb}
\langle f|g\rangle = {2h-1 \over 2\pi}\int_D  {d^2u\over  (1-u\overline{u})^{2-2h}} \overline{f(u)}g(u)~.
\eea
The action of the complexified Lie algebra $SU(1,1)$ (i.e. $SL(2,\CC)$)  on holomorphic functions is given by,
\bea
\label{xba}
L_1 = \partial_u~,\quad L_0 = u\partial_u +h~,\quad L_{-1}=u^2 \partial_u+2hu~,
\eea
with $[L_m,L_n]=(m-n)L_{m+n}$.  This inner product gives the adjoint operators as $L_n^\dagger = L_{-n}$.  The inner product $\< f |  g \>$ is invariant under the simultaneous transformations of $f$ and $g$ by $SU(1,1)$, generated by the self-adjoint combinations  $L_0, L_1+L_{-1}, i(L_1-L_{-1})$.
The perturbative $1/c$ expansion of the Wilson line requires only a representation of the $SL(2,\RR)$ Lie algebra. As such, the only ingredients required for the computation of the matrix elements considered in this paper are the commutation relations of the generators $L_m$, the inner product, and the adjoint operator relation. The resulting formulas for the matrix elements make sense for arbitrary $h>1/2$, and provide the desired analytic continuation of the spin $j$ matrix elements.

As reviewed in section \ref{Wilsec}, the Wilson line was built on $SL(2,\RR)$ states obeying
\bea\label{xaf}
&&L_{-1}|h;{\rm in}\rangle =0~,\quad L_0 |h;{\rm in}\rangle =-h|h;{\rm in}\rangle \cr
&& L_{1}|h;{\rm out}\rangle =0~,\quad L_0 |h;{\rm out}\rangle =h|h;{\rm out}\rangle~,
\eea
These states therefore correspond to the functions
\bea\label{xbc}   |h_{\rm in} \rangle \rightarrow u^{-2h}~,\quad  |h_{\rm out} \rangle \rightarrow 1~.
\eea
The computation of the Wilson line therefore maps to integrals on the disk.  For example at lowest order the two-point function is recovered as
\bea\label{xyy} W[z_2,z_1]&=&\langle h; {\rm out}| \exp \left \{ \int_{z_1}^{z_2} dz L_1 \right\} |h;{\rm in}\rangle  \cr
&= &  {2h-1 \over 2\pi}\int_D{d^2u \over (1-u\overline{u})^{2-2h}} (u+z_2-z_1)^{-2h}\cr
& = & (z_2-z_1)^{-2h}  ~.\eea
Higher order terms in the $1/c$ expansion just involve additional insertions of the $L_n$.
%
%
%
It is easy to see that order-by-order in $1/c$ this gives the same result as working with spin $j$ representations and then setting $j=-h$ at the end.

\section{$SL(2,\RR)$ matrix elements}
\setcounter{equation}{0}
\label{sec:A}

In this appendix, we derive a recursion relation for the $SL(2,\RR)$ group theory factors which enter into the calculation of the large $c$ expansion of matrix elements of the gravitational Wilson line  operator. The factors of interest are the functions $F_n$ defined by,
\bea
z^{2j} F_n (z;y_n, \cdots, y_1) = \< j,-j| e^{zL_1} X(y_n) \cdots  X(y_1) | j, j\>
\eea
where $X(y)=L_{-1} - 2 yL_0 + y^2 L_1$. Furthermore,  $|j,j\>$ denotes the highest weight state  of a representation of $SL(2,\RR)$ with finite dimension  $2 j+1 \in \NN$ and thus satisfies $L_{-1} |j,j\>=0$. Choosing unit norm for $|j,j\>$ sets $F_0(z)=1.$ To obtain a recursion relation for the matrix elements $F_n$ we recursively define the states $\mS_n$  by,
\bea
\label{sl}
\mS_{n} (y_n, \cdots, y_1) = X(y_n) \,  \mS_{n-1} (y_{n-1}, \cdots, y_1),
\hskip 1in
\mS_0 = |j,j\>,
\eea
or equivalently $\mS_n (y_n, \cdots, y_1) = X(y_n)  \cdots X(y_1)  |j,j\>$.
Commuting the operators $L_{-1}$ and $L_0$ in each $X$-factor  to the right and evaluating the result on $|j,j\>$ shows that $\mS_n$ is a linear combination of states $L_1^k |j,j\>$ with coefficients $S_n^{(k)}$,
\bea
\mS_n(y_n, \cdots, y_1)  = \sum _{k=0}^n S_n^{(k)} (y_n, \cdots, y_1) L_1^k \,  |j,j\>.
\eea
Implementing the recursion relations on the states $\mS_n$ given by (\ref{sl})
produces the following recursion relations on the coefficients $S_n ^{(k)}$,
\bea
\sum _{k=0}^{n+1} S_{n+1} ^{(k)}  L_1^k \,  |j,j\>
=
\sum _{\ell=0}^n S_n^{(\ell)}  \Big ( \ell(\ell - 2j-1) L_1^{\ell-1}  - 2 y_{n+1} (j-\ell) L_1^\ell  + y_{n+1}^2 L_1^{\ell+1} \Big )    |j,j\>.
\eea
Assuming that $j$ is large enough, namely for $n+1 < 2j$,  the states $ L_1^k|j,j\>$ for $0\leq k \leq n+1$ will all be  linearly independent. Identifying their coefficients on both sides gives the following recursion relations for $0 \leq k \leq n+1$,
\bea
S_{n+1}^{(k)} = y_{n+1}^2 \, S_n ^{(k-1)} - 2 (j-k) y_{n+1} S_n ^{(k)} +(k+1)(k-2j) S_n ^{(k+1)}
\eea
where $S_0^{(0)}=1$ and we  set $S_n ^{(k)} =0$ whenever $k <0$ or $k >n$. The truncations $S_n^{(k)}=0$ which arise for $n \geq k > 2j$, follow automatically from the recursion relations for $j$.   Finally, we derive the formula for $F_n$ in terms of $S_n ^{(k)}$ by using the matrix elements $\< j, -j| e^{zL_1} |j,j\> = z^{2j}$ and their $z$-derivatives,
\bea
F_n (z;y_n, \cdots, y_1) = \sum _{k=0}^n {\Gamma (2j+1) z^{-k} \over \Gamma (2j+1-k)}
S_n ^{(k)}(y_n, \cdots, y_1).
\eea
By construction, the combination $z^nF_n(z;y_n,\ldots,y_1)$ is a homogeneous polynomial in the variables $z,y_1 ,\ldots , y_n$ of combined degree $2n$.


\section{Gravitational Wilson line computations}
\setcounter{equation}{0}
\label{sec:B}

In this appendix we discuss the calculations of the coefficients $W_{1 \cdots n}$ and $W_P$   required to evaluate $W_\ep (z)$ in (\ref{wpar}). The numerator functions $\cN_n$ are given by (\ref{nn}) in terms of the functions $F_n$ computed  in Appendix \ref{sec:A}. They are polynomials in $\alpha _1 , \cdots \alpha _{n-1}$ with coefficients which have simple poles in $\ep$. Their expressions rapidly become lengthy as $n$ increases, and were handled by MAPLE.  The denominator functions $\cD_P$  will be listed below.

\subsection{Computation of $W_{12}$ and $W_{123}$}

The denominator functions for $n=2,3$ are given as follows,
\bea
\cD_{12}  =   1,
\hskip 1in
\cD_{123}  =   (1-\alpha _2)^{2 - \ep} \alpha _2^{2-\ep} .
\eea
The integration over $\alpha _1$ may be carried out by using the  $\delta$-function,  and we have,
\bea
W_{12}  =  { 6^2 \over 2 c}  \, \cN_2 (1),
\hskip 1in
W_{123} = { 6^3 \over c^2} \,  \int _0 ^1 d\alpha_{2} \,
{ \cN_3 (1-\a_2, \a_2) \over \alpha _2^{2-\ep} (1-\alpha _2)^{2 - \ep} }
\eea
which leads to the results on the first two lines of  (\ref{Wup4}). Since $\cN_3(1-\alpha_2, \alpha_2)$ is polynomial in $\alpha _2$, the only integrals required to evaluate $W_{123}$ are of the Euler type given in (\ref{euler}).

\subsection{Calculation of $W_{1234}$}

For $n=4$ the different partitions give the following denominator functions,
 \begin{align}
 \cD_{[12][34]} & =  \alpha _1 ^{4-2\ep}\alpha _3 ^{4-2\ep}
 & \alpha _2
 \no \\
 \cD_{[13][24]} & =  (\alpha_1+\alpha_2)^{4-2\ep}(\alpha _2 +\alpha _3)^{4-2\ep}
 & \alpha _2
 \no \\
 \cD_{[14][23]} & =  \alpha _2^{4-2\ep}
 & \alpha _3
 \no \\
 \cD_{[1234]} & =  \alpha _1^{2-\ep} \alpha _2 ^{2-\ep} \alpha _3 ^{2-\ep}
 & \alpha _3
 \no \\
 \cD_{[1324]} & =  (\alpha _1 + \alpha _2)^{2-\ep} \alpha _2 ^{2-\ep} (\alpha _2+\alpha _3)^{2 - \ep}
 &\alpha _2
 \no \\
 \cD_{[1342]} & =  \alpha _1 ^{2-\ep} (\alpha _1+\alpha _2)^{2-\ep} (\alpha _2 + \alpha _3)^{2-\ep} \alpha _3 ^{2-\ep}
 & \alpha _2
 \end{align}
where the right column lists a convenient choice of variable to be eliminated with the help of the $\delta$-function.
Since $\cN_4$ is polynomial in $\alpha_1, \alpha_2, \alpha _3$, the integrals required to evaluate $W_{[12][34]}, W_{[13][24]}$ and $W_{[14][23]}$ are of the Euler beta function type of (\ref{euler}). They may be readily evaluated and produce the results on the third line of (\ref{Wup4}).

\sm

The evaluation of $W_{[1234]}$ proceeds analogously. For $W_{[1324]}$, however, a set of non-standard  integrals is required. They are denoted by $\cK_\ep(a,b,c)$ and are calculated in Appendix \ref{sec:D2}. Similarly, for $W_{[1342]}$
another set of non-standard integrals is required which are denoted $\cJ_\ep(a,b)$ and evaluated in Appendix \ref{sec:D1}.
Here and below, the nature of these non-standard integrals is dictated by the structure of the denominator functions.

\subsection{Calculation of $W_{12345}$}

For $n=5$ the denominator functions are given by,
\begin{align}
\cD_{[12][345]} & =  \a_1^{4-2\ep} \a_3^{2-\ep}  (\a_3+\a_4)^{2-\ep} \a_4^{2-\ep}
& \a_2
\no \\
\cD_{[13][245]} & =  (\a_1+\a_2) ^{4-2\ep} (\a_2+\a_3)^{2-\ep}  (\a_2+\a_3+\a_4)^{2-\ep} \a_4^{2-\ep}
& \a_3
\no \\
\cD_{[14][235]} & =  (\a_1+\a_2+\a_3) ^{4-2\ep} \a_2^{2-\ep}  (\a_2+\a_3+\a_4)^{2-\ep} (\a_3+\a_4)^{2-\ep}
& \a_3
\no \\
\cD_{[15][234]} & =   \a_2^{2-\ep} \a_3^{2-\ep} (\a_2+\a_3)^{2-\ep}
& \a_4
\no \\
\cD_{[23][145]} & =   (\a_1+ \a_2+\a_3)^{2-\ep}   \a_2^{4-2\ep} \a_4^{2-\ep}
& \a_3
\no \\
\cD_{[24][135]} & =   (\a_1+ \a_2)^{2-\ep}  ( \a_2+\a_3) ^{4-2\ep} (\a_3+\a_4)^{2-\ep}
& \a_4
\no \\
\cD_{[25][134]} & =   (\a_1+ \a_2)^{2-\ep} (\a_1+ \a_2+\a_3)^{2-\ep}  ( \a_2+\a_3+\a_4) ^{4-2\ep} \a_3^{2-\ep}
& \a_2
\no \\
\cD_{[34][125]} & =   \a_1^{2-\ep} ( \a_2+\a_3+\a_4)^{2-\ep}   \a_3^{4-2\ep}
& \a_4
\no \\
\cD_{[35][124]} & =   \a_1^{2-\ep} (\a_1+ \a_2+\a_3)^{2-\ep}  (\a_2+\a_3)^{2-\ep} (\a_3+\a_4)^{4-2\ep}
& \a_2
\no \\
\cD_{[45][123]} & =   \a_1^{2-\ep} (\a_1+\a_2)^{2-\ep} \a_2^{2-\ep} \a_4^{4-2\ep}
& \a_3
\end{align}
The integrals required for the coefficients $W_{[12][345]}, W_{[15][234]}, W_{[23][145]}, W_{[34][125]}, W_{[34][125]}$, and $W_{[45][123]}$ may be reduced to integrals of the Euler type in (\ref{euler}) using judicious choices of variables. For example, in  $W_{[45][123]} $  we integrate over $\a_3$,  keep the variable $\a_4$,  and change variable from $\a_1,\a_2$ to $t, \beta$ with $\a_1 =   (1-\a_4) t  \beta$ and $\a_2 = (1-\a_4) t (1-\beta)$,  so that $0 \leq t, \beta \leq 1$. In terms of these variables, and letting $\alpha _4 \to 1-\a_4$,   the integral becomes,
\bea
c^3 \, W_{[45][123]}  =
{6^5 \over 2} \int _0^1 d\a_4 \int _0^1  dt  \int _0^1 d \beta \,
{ \cN_5(\alpha _4 t\beta , \a_4 t (1-\beta)  ,  \a_4(1-t), 1- \a_4)
\over \a_4^{4-3\ep} (1-\a_4)^{4-2\ep}  \, t^{5-3\ep} \, \beta ^{2-\ep} (1-\beta)^{2-\ep} }~.
\eea
To evaluate the decoupled integrals we expand the numerator $\cN_5$ into powers of $\alpha_4, t,  \beta$,
\bea
\cN_5(\alpha _4 t\beta , \a_4 t (1-\beta)  ,  \a_4(1-t), 1- \a_4) =
\sum_{A=0}^4 \sum_{B=0}^6 \sum _{C=0}^2  \alpha _4^A \, t^B \, \beta ^C \cM_{A,B,C}
\eea
and use,
\bea
&&
\int _0 ^1 d \alpha _4 \int _0 ^1 dt \int _0 ^1 d \beta
{\alpha _4 ^A \, t^B \, \beta ^C \over  \a_4^{4-3\ep} (1-\a_4)^{4-2\ep} \, t^{5-3\ep} \, \beta ^{2-\ep} (1-\beta)^{2-\ep}}
\no \\ && \hskip 1in
=
 { \Gamma (A-3+3\ep) \Gamma (-3+2\ep)  \Gamma (C-1+\ep) \Gamma (-1+\ep) \over (B-4+3\ep) \Gamma (A-6+5\ep) \Gamma (C-2+2\ep)}~.
\eea

\sm
The integrals for the remaining partitions $W_{[13][245]}$,  $W_{[35][124]}$, $W_{[14][235]}$, $W_{[25][134]}$ are closely related to one another. They may be evaluated   in terms of nested integrals $\cL_\ep (a,b,c,f)$ computed in the Appendix \ref{sec:D3}. For example, in $W_{[13][245]}$ we integrate over $\alpha_3$ with the help of the $\delta$-function, and
change variables from $\alpha _2 $ to $\beta = \alpha _1+\alpha_2$,
\bea
W_{[13][245]}= {6^5 \over 2} \int _0 ^1 d \alpha _1 \int _0 ^{1-\alpha_1} d\alpha _4 \int _{\alpha_1}  ^{1 -\alpha_4} d \beta \,
{ \cN_5 ( \alpha_1, \beta - \alpha _1, 1-\beta-\alpha _4, \alpha _4)
	\over \beta^{4-2\ep} (1-\alpha _1)^{2-\ep}  (1-\alpha _1 -\alpha _4)^{2-\ep} \alpha _4 ^{2 - \ep}}~.
\eea
The polynomial $\cN_5$ is a quadratic in each variable $\alpha_1, \alpha _4, \beta$.  Expanding in powers of $\beta$, for fixed $\alpha_1, \alpha_4$,  we obtain,
\bea
{6^5 \over 2} \cN_5 ( \alpha_1, \beta - \alpha _1, 1-\beta-\alpha _4, \alpha _4)= \sum _{B=0}^2 \beta ^B \cM_B (\alpha_1, \alpha_4)
\eea
where the functions $\cM_B(\alpha_1, \alpha_4)$ are quadratic polynomials in $\alpha _1$ and $\alpha_4$.
The integral over $\beta$ may now be performed term by term in powers of $\beta$,
\bea
W_{[13][245]}  = \sum _{B=0}^2 {W^{(4)}_B - W^{(1)}_B \over B-3+2\ep}
\eea
where,
\bea
W^{(1)}_B & = &  \int _0 ^1 d \alpha _1 \int _0 ^{1-\alpha_1} d\a_4 \,
{ \cM_B ( \alpha_1,  \alpha _4)
	\over \alpha _1 ^{3-B-2\ep} (1-\alpha _1)^{2-\ep}  (1-\alpha _1 -\alpha _4)^{2-\ep} \alpha _4 ^{2 - \ep}}~,
\no \\
W^{(4)}_B & = &  \int _0 ^1 d \alpha _1 \int _0 ^{1-\alpha_1} d\a_4 \,
{ \cM_B ( \alpha_1,  \alpha _4)
	\over  (1-\alpha _1)^{2-\ep}  (1-\alpha _1 -\alpha _4)^{2-\ep} \alpha _4 ^{2 - \ep} (1-\alpha_4)^{3-B-2\ep}}~.
\eea
In the integral for $W_B^{(1)}$, we decouple the integrations by changing variables from $\alpha _4$ to $\alpha _4=(1-\alpha_1)t$, and then perform the integrations using (\ref{euler}). The evaluation of $W^{(4)}_B$ is considerably more complicated. We expand $\cM_B$ is powers of $(1-\alpha_1)$ and $(1-\alpha _4)$,
\bea
\cM_B (\alpha _1, \alpha _4) & = & \sum _{A=0}^2 \, \sum _{C=0}^2 (1-\alpha _1)^A \, (1-\alpha _4)^C \, \cM_{A,B,C}^{(4)}
\no \\
W^{(4)}_B  & = &  \sum _{A,C=0}^2  \cM_{A,B,C}^{(4)} \, \cL_\ep  (A-1, -1, B+C-2, -1)
\eea
where the family of integrals $\cL_\ep (a,b,c,f)$ is defined and evaluated in Appendix \ref{sec:D3}.

\subsection{Calculation of $W_{123456}$}

Finally, the denominator functions for $n=6$ are given by,
\begin{align}
\cD_{[12][34[56]} & =  \a_1^{4-2\ep} \a_3^{4-2\ep}   \a_5^{4-2\ep}
& \a_4
\no \\
\cD_{[12][35[46]} & =  \a_1^{4-2\ep} (\a_3+\a_4)^{4-2\ep}  (\a_4+ \a_5)^{4-2\ep}
& \a_2
\no \\
\cD_{[12][36[45]} & =  \a_1^{4-2\ep} (1-\a_1-\a_2)^{4-2\ep}  \a_4^{4-2\ep}
& \a_5
\no \\
\cD_{[13][24[36]} & =  (\a_1+\a_2)^{4-2\ep} (\a_2+\a_3)^{4-2\ep}  \a_5^{4-2\ep}
& \a_4
\no \\
\cD_{[13][25[46]} & =  (\a_1+\a_2)^{4-2\ep} (1-\a_1-\a_5)^{4-2\ep} (\a_4+ \a_5)^{4-2\ep}
& \a_3
\no \\
\cD_{[13][26[45]} & =  (\a_1+\a_2)^{4-2\ep} (1-\a_1)^{4-2\ep} \a_4^{4-2\ep}
& \a_5
\no \\
\cD_{[14][23[56]} & =  (\a_1+\a_2+\a_3)^{4-2\ep} \a_2^{4-2\ep} \a_5^{4-2\ep}
& \a_4
\no \\
\cD_{[14][25[36]} & =  (\a_1+\a_2+\a_3)^{4-2\ep} (1-\a_1-\a_5) ^{4-2\ep} (\a_3+\a_4+\a_5)^{4-2\ep}
& \a_3
\no \\
\cD_{[14][26[35]} & =  (\a_1+\a_2+\a_3)^{4-2\ep} (\a_3+\a_4) ^{4-2\ep} (1-\a_1)^{4-2\ep}
& \a_5
\no \\
\cD_{[15][23[46]} & =  (1-\a_5)^{4-2\ep} \a_2 ^{4-2\ep} (\a_4+\a_5)^{4-2\ep}
& \a_3
\no \\
\cD_{[15][24[36]} & =  (1-\a_5)^{4-2\ep} (\a_2+\a_3) ^{4-2\ep} (1-\a_1-\a_2)^{4-2\ep}
& \a_4
\no \\
\cD_{[15][26[34]} & =  (1-\a_5)^{4-2\ep} \a_3 ^{4-2\ep} (1-\a_1)^{4-2\ep}
& \a_4
\no \\
\cD_{[16][23[45]} & =  \a_2^{4-2\ep} \a_4 ^{4-2\ep}
& \a_5
\no \\
\cD_{[16][24[35]} & =  (\a_2+\a_3)^{4-2\ep} (\a_3+\a_4) ^{4-2\ep}
& \a_5
\no \\
\cD_{[16][25[34]} & =  (\a_2+\a_3+\a_4)^{4-2\ep} \a_3 ^{4-2\ep}
& \a_5
\end{align}
The integrals required for the coefficients  $W_{[12][34][56]}$,  $W_{[12][35][46]}=W_{[13][24][56]}$,   $W_{[12][36][25]}$,
$W_{[13][25][46]}$, $W_{[13][26][45]}$, $W_{[14][23][56]}$, $W_{[15][23][46]}$, $W_{[16][23][45]}$, $W_{[16][24][35]}$,
$W_{[16][25][34]}$ may be evaluated using judicious variables and the Euler formula of (\ref{euler}). The integrals required for the coefficients $W_{[14][25][36]}$, $W_{[14][26][35]}$, $W_{[15][24][36]}$, $W_{[15][26][34]}$ may be evaluated using the family of integrals with $\cK_{2\ep}$ evaluated in Appendix \ref{sec:D}. Putting everything together we get the result reported in the first line of (\ref{Wup6}).


\section{Calculation of $\<T_\infty W_\ep(z)\>$}
\setcounter{equation}{0}
\label{sec:C}

The calculation of $\<T_\infty W_\ep(z)\>$ is parallel to the calculation of $\<W_\ep (z)\>$ already given. The expansion of the path ordered exponential (\ref{TW}) may be organized as follows,
\bea
\< T_\infty W(z) \> = z^{2j+2} \sum _{n=0}^\infty \alpha ^n \, z^{(n-1) \ep} \, TW_{x 1\cdots n}(z)
\eea
where the coefficients $TW_{x 1 \cdots n } $ are independent of $z$ and given by,
\bea
\label{TWF}
TW_{x 1\cdots n}(z) = { 6^n \over  c^n }
\int _0 ^1 dy_n  \cdots \int ^{y_2}_0 dy_1 F_n(1;y_n, \cdots, y_1)
\< T_\infty T^n \> _{x 1 \cdots n}
\eea
where we use the following notation,
\bea
\< T_\infty T^n \> _{x 1 \cdots n} = \lim _{x \to \infty} \Big ( x^{4-2\ep} \< T(x) T(y_1) \cdots T(y_n)  \> \Big ).
\eea
 The symbol $x$ used in the subscript to $TW_{x1\cdots n}$ and $\< T_\infty T^n \> _{x 1 \cdots n} $ stands for a place-holder indicating the position of the operator $T(x)$ in the correlator.

 \sm

 The stress tensor correlators are evaluated using the same decomposition into partitions of one-loop cycles that we have used for the calculation of $W_{1\cdots n}$, and the relevant correlators are given as follows. Evidently we have $ \< T_\infty T^0 \> _{x} = 0$ and  $  \< T_\infty T^1 \> _{x 1} = { c \over 2}$, as well as the following formula for cycles of arbitrary length $n+1$,
\bea
\< T_\infty T^n\>_{[x1 \cdots n]} = { c \over |y_1-y_2|^{2-\ep}  |y_2-y_3|^{2-\ep} \cdots |y_{n-1} - y_n |^{2 - \ep}}~.
\eea
The correlators we need are as follows,
\bea
 \< T_\infty T^3 \> _{x 123} & = &  \< T_\infty T^3 \> _{[x 1] [23]} +  \< T_\infty T^3 \> _{[x 2] [31]}
 +  \< T_\infty T^3 \> _{[x 3] [12]}
 \no \\ &&
 +  \< T_\infty T^3 \> _{[x 123]}  +  \< T_\infty T^3 \> _{[x 132]} +  \< T_\infty T^3 \> _{[x 213]} ,
\no \\
 \< T_\infty T^4 \> _{x 1235} & = &
  \< T_\infty T^4 \> _{[x1][234]} + \hbox{ 3 more partitions}
  \no \\ &&
  +  \< T_\infty T^4 \> _{[x12][34]} + \hbox{ 5 more partitions},
\no \\
 \< T_\infty T^5 \> _{x 12345} & = &
  \< T_\infty T^5 \> _{[x 1][23][45]} + \hbox{ 14 more partitions}.
\eea
The contribution from a partition $P$ is given by the product of the contributions of all the cycles in the partition, just as in  (\ref{TC}), but including now also the point $x$.

\sm

The integrals in (\ref{TWF}) may again be simplified with the help of the change of variables used for $W$ in (\ref{yau}), and we obtain the following final formula,
\bea
\label{TWP}
TW_P = { 6^n \over 2^{p_2} c^{n-p}}   \int _0 ^1 d\alpha_{n-1} \cdots \int ^1_0 d\alpha _1 \,
\delta \left (1-\sum_{k=1}^{n-1} \alpha_k \right )
{ \cT\cN_n (\alpha_1, \cdots, \alpha _{n-1}) \over \cT \cD_P (\alpha_1, \cdots, \alpha _{n-1})}
\eea
where $p$ is the total number of cycles in the partition $P$ and $p_2$ is the number of 2-cycles.
The function $\cT \cN_n$ is given by,
\bea
\cT\cN_n (\alpha _1, \cdots , \alpha _{n-1}) =  \int _0 ^1 du \, u^{-n+ (n-1) \ep} \int _u ^1 dx_n \,
   F_n(1;y_n, y_{n-1}, \cdots, y_1).
 \eea
 Note that the integrand of $\cT \cN_n$  differs  in the variable $u$ from the one for $\cN_n$ used in the calculation of $W$. The function $\cT\cD_P$ is given in terms of the stress tensor correlators by,
\bea
\< T_\infty T^n \>_P = { c^p \over 2^{p_2}} { u^{-(2 -  \ep)(n-1)}  \over \cT \cD _P (\alpha _1, \cdots, \alpha _{n-1})}~.
\eea
One of the $\alpha _k$-integrals may be carried out by satisfying the $\delta$-function, so that the number of non-trivial integrals left over is $n-2$.

\subsection{Calculation of $TW_{x1\cdots n}$}

Since $TW_{x}$ involves the expectation value of a single stress tensor, it vanishes. One also readily shows that $TW_{x1} =-j$. For higher values of $n$, the expressions for $\cT\cN_n$ rapidly become lengthy and the corresponding calculations have been carried out using MAPLE. The $\alpha$-integrals involved are less exotic than the ones that were needed for the calculation of $W_\ep(z)$, and may easily be worked out. To orders $1/c$ and $\ep$ the coefficients are given by,
\bea
c \, TW_{x12} & = & { 12 j \over \ep} -  j(18 j+13) + { j \over 12} (162 j + 259) \ep,
\no \\
c \, TW_{x123} ^{(1)} & = &  -{6j(j^2+j+1) \over \ep} -2 j (5j^2-4j-5) - { j \over 3}  (74 j^2 + 128 j+107) \ep .
\eea
To order $1/c^2$ we have the following contributions,
\bea
c^2 \, TW_{x123} ^{(2)}  & = &
- { 144 j \over  \ep^2}  + {48 j \over  \ep } (9j+8)
- { 48 \pi^2 j \over 5}  (j^2+j-2) - 8j (36 j^2+144j+131),
\no \\
c^2 \, TW_{x1234} & = &
{24 j \over \ep^2} (7j^2+7j+8) -{ 2 j \over \ep} (78j^3-75j^2+279 j + 274)
\no \\ &&
 - { j \over 6} ( 1242 j^3 -10863 j^2 -21717 j -15880) -16 \pi^2 j (j^2+j+4),
 \no \\
 c^2 \, TW_{x12345} & = &
- {18 j \over \ep^2} (j^2+j+1)(j^2+j+3)
-{ 3 j \over \ep} (20 j^4 -8 j^3 + 25 j^2 -55 j -59)
\no \\ &&
 - 2 j ( 99 j^4 +102 j^3 + 753 j^2 + 1065 j + 622) + 24 \pi^2  j (j^2+j+2)
\eea
where $TW_{x123}=TW_{x123}^{(1)} +TW_{x123}^{(2)} $.


\section{Non-standard integrals}
\setcounter{equation}{0}
\label{sec:D}

The most basic integral we use throughout is Euler's beta function formula,
\bea
\label{euler}
\int _0^1 d \alpha \, \alpha ^{s-1} (1-\alpha )^{t-1} = {\Gamma (s) \Gamma (t) \over \Gamma  (s+t)}~.
\eea
Next, we  evaluate various  non-standard integrals, needed in an expansion in powers of $\ep$.

\subsection{The $\cJ_\ep (A,B)$ integrals}
\label{sec:D1}

The  integrals are defined by,
\bea
\cJ_\ep (A,B) = \int _0^1 d \alpha \int _0 ^{1-\alpha} d \beta \, { \alpha ^A \beta ^B \over \alpha ^{2-\ep} (1-\alpha)^{2-\ep} \beta ^{2-\ep} (1-\beta)^{2-\ep}}
\eea
for  integers $A,B$ in the range $0 \leq A,B\leq 2$. In view of the symmetry of the integration under the interchange of $\alpha$ and $\beta$, we have $\cJ_\ep (A,B)= \cJ_\ep (B,A)$, reducing the number of integrals needed from 9 to 6. We begin by evaluating the following auxiliary integrals,
\bea
I_{a,b} (s,t) = \int _0^1 d \alpha \int _0 ^{1-\alpha} d \beta (1-2\alpha)^a (1-2\beta)^b \alpha ^{s-1} (1-\alpha )^{s-1} \beta ^{t-1} (1-\beta)^{t-1}
\eea
for positive integers $a,b$. Clearly, we have $I_{a,b}(s,t)=I_{b,a}(t,s)$ and the integrals $J_\ep(A,B)$ are linear combinations of the integrals $I_{a,b}(\ep-1,\ep-1)$ for various values of $a,b$. In view of the identity $(1-2\alpha)^2 = 1-4 \alpha (1-\alpha)$ and its analogue for $\beta$, we have the following relations,
\bea
I_{a+2,b}(s,t) & = & I_{a,b}(s,t) - 4 I_{a,b}(s+1,t),
\no \\
I_{a,b+2}(s,t) & = & I_{a,b}(s,t) - 4 I_{a,b}(s,t+1)
\eea
allowing us to restrict the range to $0\leq a,b \leq 1$.  In view of these symmetries and relations, the remaining  integrals  may be evaluated using (\ref{euler}), and we have $I_{1,1}(s,t)=0$ as well as,
\bea
I_{0,0} (s,t) = { \Gamma (s)^2 \, \Gamma (t)^2 \over 2 \, \Gamma (2s) \, \Gamma (2t)}~,
\hskip 0.6in
I_{0,1}(s,t) = { \Gamma (s+t)^2 \over t \, \Gamma (2s+2t)}~.
\eea
Explicit expressions for the required $J_\ep (A,B)$ in terms of $I_{a,b}(s,t)$ are given as follows,
\bea
\cJ_\ep (0,0) & = & I_{0,0} (\ep-1,\ep-1),
\no \\
\cJ_\ep (1,0) & = & - \half I_{1,0} (\ep-1,\ep-1) + \half \cJ_\ep (0,0),
\no \\
\cJ_\ep (2,0) & = & -  I_{0,0} (\ep,\ep-1) + \cJ_\ep (1,0) ,
\no \\
\cJ_\ep (1,1) & = & \cJ_\ep (1,0) - {1 \over 4} \cJ_\ep (0,0),
\no \\
\cJ_\ep (2,1) & = & \half I_{0,1}(\ep,\ep-1) + \cJ_\ep (1,1) + \half \cJ_\ep (2,0) - \half \cJ_\ep (1,0),
\no \\
\cJ_\ep (2,2) & = & \cJ_{\ep+1}  (0,0) + 2 \cJ_\ep (2,1) - \cJ_\ep (1,1).
\eea

\subsection{The $\cK_\ep (a,b,c)$ integrals}
\label{sec:D2}

We shall also need integrals of the following form,
\bea
\cK_\ep (a,b,c) = \int _0 ^1 d \alpha \int _0 ^{1-\alpha} d \beta \,
(1-\alpha)^{a-1+\ep} (1-\beta) ^{b-1+\ep} (1-\alpha - \beta)^{c-1+\ep}.
\eea
for several sets of integers $a,b,c$. Clearly we have $\cK_\ep ( a,b,c) = \cK_\ep (b,a,c)$. We use the identity $(1-\alpha)+(1-\beta)-(1-\alpha-\beta)=1,$ and integration by parts in $\alpha$ and in $\beta$ to find the following formulas,
\bea
(a+b+c+3\ep )  \cK_\ep (a+1,b,c)  & = & (a+\ep) \cK_\ep (a,b,c) + { 1 \over a+c+2\ep}~,
\\
(a+b+c+3\ep ) \cK_\ep (a,b+1,c)  & = & (b+\ep) \cK_\ep (a,b,c) + { 1 \over b+c+2\ep}~,
\no \\
(a+b+c+3\ep ) \cK_\ep (a,b,c+1)  & = & - (c+\ep) \cK_\ep (a,b,c) + { 1 \over a+c+2\ep} + { 1 \over b+c+2\ep}~.
\no
\eea
To initialize the recursion relations in all three integers $a,b,c$ it suffices to compute $\cK_\ep (a,b,c)$ at a point in the domain of the variables $a,b,c,\ep$ where it is given by a convergent integral. For example, $\cK_\ep (1,1,1)$ is given by an absolutely convergent integral for $-1 < \Re(\ep) $, and admits a convergent Taylor expansion in $\ep$ around $\ep=0$. To order $\ep^2$, it is given as follows,
\bea
\cK_\ep (1,1,1) = \half -{5 \over 4} \ep + { 11 \over 8} \ep^2 + { \pi^2 \over 12} \ep^2 + \cO(\ep^3).
\eea
The expressions for $\cK_\ep (a,b,c)$ for the values $-1 \leq a,b \leq 1$ and $c=-1$ needed for the evaluation of $W_{[1324]}$ are obtained using the recursion relations through MAPLE.

\subsection{The $\cL_\ep (a,b,c,f)$ integrals}
\label{sec:D3}

The integrals are defined by,
\bea
\cL_\ep (a,b,c,f)=
\int _0 ^1  d \alpha  \int _0 ^{1-\alpha}  d\beta \,
(1-\alpha )^{a-1+\ep}   \beta ^{b-1+\ep} (1-\beta)^{c-1+2\ep } (1-\alpha  -\beta)^{f-1+\ep}
\eea
for integer values of $a,b,c,f$. A first pair of recursion relations on the indices $a,b,c,f$ is obtained by inserting the identities $(1-\alpha)+(1-\beta) -(1-\alpha -\beta) =1$ and $\beta + (1-\beta)=1$ into the integrand, while a second set is obtained by evaluating the $\alpha$ and $\beta$-derivatives of the integrand, in each case expressing the result in terms of $\cL_\ep $-functions. We may solve  this linear system  to obtain four one-step recursion relations, given by,
\bea
Z \cL_\ep (a,b+1,c,f) & = & (Z-c-2\ep ) \cL_\ep (a,b,c,f) - R,
\\
Z \cL_\ep (a,b,c+1,f) & = & (c+2\ep ) \cL_\ep (a,b,c,f) + R,
\no \\
(a+f+2\ep ) Z \cL_\ep (a+1,b,c,f) & = &  (a+\ep )(Z-c-2\ep ) \cL_\ep (a,b,c,f) + (Z-a-\ep ) R,
\no \\
(a+f+2\ep ) Z \cL_\ep (a,b,c,f+1) & = & - (f+\ep )(Z-c-2\ep) \cL_\ep (a,b,c,f) + (Z+f+\ep ) R.
\no
\eea
where we have used the following abbreviations,
\bea
Z= a+b+c+f -1+ 5 \ep,
\hskip 1in
R =  { \Gamma (b+\ep ) \Gamma (c+f+3\ep ) \over \Gamma (b+c+f+4\ep)}~.
\eea
The recursion relations may be initialized by the absolutely convergent integral  $\cL_\ep (1,1,1,1)$ for $\ep $ near 0,  in an expansion in powers of $\ep$,
\bea
\cL_\ep (1,1,1,1) = \half -{9 \over 4} \ep -{ \pi^2 \ep^2 \over 12} +{53 \over 8} \ep^2 + \cO(\ep^3).
\eea
The expressions for $\cL_\ep (a,b,c,f)$ for the other required values of $a,b,c,f$ are obtained using the recursion relations through MAPLE.

\subsection{Evaluating the integrals $\cQ^{(i)} (a,b)$ for $i=1,2,3$}
\label{sec:D4}

The integrals are defined as follows,
\bea
\label{Q123}
\cQ^{(1)}_\ep (a,b) & = & \int _0 ^1 d \alpha \int ^{1-\alpha} _0 d \beta \,
\alpha ^{a-1+\ep} \beta ^{b-1+\ep} (1-\a)^\ep (1-\b )^\ep (1-\a-\b)^\ep,
\no \\
\cQ^{(2)}_\ep (a,b) & = & \int _0 ^1 d \alpha \int ^{1-\alpha} _0 d \beta \,
\alpha ^{\ep} \beta ^{\ep} (1-\a)^{a-1+\ep} (1-\b )^{b-1+\ep} (1-\a-\b)^\ep,
\no \\
\cQ^{(3)}_\ep (a,b) & = & \int _0 ^1 d \alpha \int ^{1-\alpha} _0 d \b \,
\alpha ^{\ep} (1-\a)^{a-1+\ep} \b ^{\ep} (1-\b)^\ep (1-\a-\b )^{b-1+\ep}.~~~~~~~~
\eea
The integrals are absolutely convergent for $\ep > -1$ and $\Re(a), \Re(b) > -\ep$.
 We shall be interested in evaluating these integrals in a small neighborhood of $\ep=0$, where they are absolutely convergent for $\Re(a), \Re(b) >0$. Beyond their ranges of convergence, the integrals need to be analytically continued.

\subsubsection{Recursion relations for $\cQ_\ep ^{(1)} (a,b)$ and $\cQ_\ep ^{(2)} (a,b)$}

The integrals $\cQ^{(1)} (a,b)$ and $\cQ^{(2)} (a,b)$ satisfy the symmetry relation,
\bea
\cQ_\ep ^{(i)} (b,a) = \cQ_\ep ^{(i)} (a,b)
\hskip 1in i=1,2.
\eea
To obtain recursion relations for $\cQ^{(1)} (a,b)$ we consider the following identity,
\bea
 \int _0 ^1 d \alpha \int ^{1-\alpha}_0 d \beta \, { \p \over \p \a} \Big (
\alpha ^{a+\ep} \beta ^{b-1+\ep} (1-\a)^{1+\ep} (1-\b )^\ep (1-\a-\b)^{1+\ep} \Big )  =0,
\eea
and its $\beta$-derivative counterpart, and express the individual contributions in terms of $\cQ^{(1)} (a,b)$.
As its turns out, $\cQ^{(2)} (a,b)$ satisfies the same recursion relations, and we have for $i=1,2$,
\bea
(a+\ep) \cQ_\ep ^{(i)} (a,b) &=& (2a+2+4\ep) \cQ_\ep ^{(i)} (a+1,b) + ( a+\ep) \cQ_\ep ^{(i)} (a,b+1)
\\ &&
- (a+2+3\ep) \cQ_\ep ^{(i)} (a+2,b) - (a+1+2\ep) \cQ_\ep ^{(i)} (a+1,b+1),
\no \\
(b+\ep) \cQ_\ep ^{(i)} (a,b) &=& (2b+2+4\ep) \cQ_\ep ^{(i)} (a,b+1) + ( b+\ep) \cQ_\ep ^{(i)} (a+1,b)
\no \\ &&
- (b+2+3\ep) \cQ_\ep ^{(i)} (a,b+2) - (b+1+2\ep) \cQ_\ep ^{(i)} (a+1,b+1).
\quad
\no
\eea
The integrals we need  (in a short series expansion in $\ep$) are for $a,b \geq -3$. The above recursion relations do not proceed by single-steps, and are considerably more complicated than those for the earlier integrals. In particular, they cannot be initialized at a single pair $(a,b)$. Instead, the above recursion relations allow us to express $\cQ_\ep ^{(i)} (a,b)$ for integer $a, b \geq -3$ as a  linear combination of $\cQ_\ep ^{(i)} (a,b)$ with $a \geq -3$ and  $b \geq 1$.  These relations are relatively involved and were handled with MAPLE.

\subsubsection{Recursion relation for $\cQ_\ep ^{(3)} (a,b)$}

Contrarily to $\cQ_\ep ^{(1)} (a,b)$ and $\cQ_\ep ^{(2)} (a,b)$, the function $\cQ_\ep ^{(3)} (a,b)$ is not symmetric in its arguments $a,b$. By expressing the vanishing of the integral over  partial derivatives with respect to $\alpha$ and $\beta$  in terms of $\cQ_\ep ^{(3)} (a,b)$, we obtain two recursion relations,
\bea
0 & = & (a+1+2\ep) \cQ_\ep ^{(3)} (a+1,b+1)  - (a+\ep) \cQ_\ep ^{(3)} (a,b+1)
\no \\ &&
+ (b+\ep ) \cQ_\ep ^{(3)} (a+2,b) - (b+\ep) \cQ_\ep ^{(3)} (a+1,b),
\no \\
0 & = & (2b+2+4\ep) \cQ_\ep ^{(3)} (a+1,b+1) - (b+1+2 \ep) \cQ_\ep ^{(3)} (a,b+1)
\no \\ &&
 - (b+2+3\ep) \cQ_\ep ^{(3)} (a,b+2)
\no \\ &&
- (b+\ep ) \cQ_\ep ^{(3)} (a+2 ,b)+ (b+\ep) \cQ_\ep ^{(3)} (a+1,b).
\eea
The last lines of both equations are the only terms  whose second argument is $b$. Adding the equations  eliminates those terms. Shifting the resulting equation by $b+1 \to b$, shifting the first equation by $a+1 \to a$, and eliminating $\cQ_\ep ^{(3)} (a+1,b)$ we obtain a formula for $\cQ_\ep ^{(3)} (a,b)$ in terms of functions with second argument $b+1$,
and thus a recursion relation in $b$,
\bea
&&
\cQ_\ep ^{(3)} (a,b) = \cQ_\ep ^{(3)} (a,b+1)
+ {  a+2b+1+6\ep \over (b+\ep) (b+1+3\ep)  }
\Big ( (a+2\ep) \cQ_\ep ^{(3)} (a,b+1)
\no \\ &&
\hskip 3.1in
- (a-1+\ep) \cQ_\ep ^{(3)} (a-1,b+1)  \Big ).
\qquad
\eea
Applying this recursion relation, the required quantities $\cQ_\ep ^{(3)} (a,b)$, for $b=-3, -2, -1, 0$ may be obtained from $\cQ_\ep ^{(3)} (a,1)$, which we  evaluate by convergent series.

\subsubsection{Initializing $\cQ_\ep ^{(1)} (a,b)$}

To evaluate the integrals $\cQ_\ep ^{(1)} (a,b)$ for $a \geq -3$ and $b \geq 1$ near $\ep=0$,  we change variables from $\alpha $ to $t$ by setting $\alpha = (1-\beta) t$ for $0 \leq t \leq 1$,  expand the factor $(1-(1-\beta)t )^\ep$ in powers of $(1-\beta)t$, and use the Euler relation (\ref{euler})  to evaluate the decoupled  integrals over $\beta, t$. It will be convenient to recast the result as follows,
\bea
\cQ_\ep ^{(1)} (a,b)
& = &
 \sum_{k=0} ^\ma { \Gamma (k-\ep) \over \Gamma (-\ep) \, k!} \,
{ \Gamma (b+\ep) \Gamma (k+a+1+3\ep) \over \Gamma (k+a+b+1+4\ep)} \,
{\Gamma (1+\ep) \Gamma (k+a+\ep) \over \Gamma (k+a+1+2\ep)}
\\ &&
- \ep  \sum_{k=\ma +1 } ^\infty
{ \Gamma (k-\ep) \Gamma (b+\ep) \Gamma (1+\ep) \Gamma (k+a+1+3\ep) \Gamma (k+a+\ep) \over
k! \, \Gamma (1 -\ep) \Gamma (k+a+b+1+4\ep) \Gamma (k+a+1+2\ep) }
\quad
\no
\eea
where $\ma = \max (0, -a)$.  The finite sum  is readily expanded in powers of $\ep$. The summand of the infinite series  grows as $k^{-2-b-3\ep}$ for large $k$. Therefore the series converges absolutely and uniformly in $\ep$  for $b+ 3 \, \Re(\ep) > -1$ which allows for $\Re(\ep) > -2/3$ in view of the assumption $b \geq 1$. The region of convergence includes the neighborhood of $\ep=0$ needed here, so that the expansion of  $\cQ_\ep ^{(1)} (a,b)$ is obtained by expanding the series term by term.

\subsubsection{Initializing $\cQ_\ep ^{(2)} (a,b)$}

The expansion for $\cQ_\ep ^{(2)} (a,b)$ for $b \geq 1$ may be obtained by the same methods and  is similar, but not identical, to the one for $\cQ_\ep ^{(1)} (a,b)$. Starting with its definition in (\ref{Q123}), we change variables from $\beta$ to $t$ with $\beta = (1-\alpha) t$ for $0 \leq t \leq 1$, expand the factor $(1-t(1-\alpha) )^{b-1+\ep}$ in powers of $t(1-\alpha)$, and perform the decoupled integrals using Euler's formula. It will be convenient to recast the result as follows,
\bea
\cQ_\ep ^{(2)} (a,b) & = &
\sum_{k=0}^\mb  {\Gamma (k-b+1-\ep) \over \Gamma(-b+1-\ep) \, k!} \,
{\Gamma(1+\ep)^2 \Gamma (k+a+1+3\ep) \Gamma (k+1+\ep) \over \Gamma (k+a+2+4\ep) \Gamma (k+2+2\ep)}
 \\ && +
\sum_{k=\mb+1}^\infty  {\Gamma (k-b+1-\ep) \over \Gamma(-b+1-\ep) \, k!} \,
{\Gamma(1+\ep)^2 \Gamma (k+a+1+3\ep) \Gamma (k+1+\ep) \over \Gamma (k+a+2+4\ep) \Gamma (k+2+2\ep)}
\quad
\no
\eea
where $\mb = \max(b-1,-a-1)$. The summand of the infinite series grows as $k^{-2-b-3\ep}$ for large $k$ and therefore the series converges absolutely and uniformly in $\ep$ for $b\geq 1$ and $\ep $ near zero. The expansion of $\cQ_\ep ^{(2)} (a,b)$ in powers of $\ep$ is obtained as it was for $\cQ_\ep ^{(1)} (a,b)$.

\subsubsection{Initializing  $\cQ_\ep ^{(3)} (a,b)$}

The recursion relation for $\cQ_\ep ^{(3)} (a,b)$ is initialized by the value of the integrals $\cQ_\ep ^{(3)} (a,1)$.
To evaluate it, we change variables  from $\beta$ to $t$ with  $\beta = (1-\a) t$ for $0\leq t \leq 1$,  expand the factor
$(1-t(1-\alpha))^\ep$  in powers of $t (1-\a)$, and perform the integrals using Euler's formula. The result is conveniently presented as follows,
\bea
\cQ_\ep ^{(3)} (a,1)  & = &
\sum _{k=0}^{\ma'} { \Gamma (k-\ep) \over \Gamma (-\ep) \, k!}
{\Gamma (1+\ep)^2 \, \Gamma (k+a+1+3\ep) \Gamma (k+1+\ep)
\over \Gamma (k+a+2+4\ep) \Gamma (k+2+2\ep)}
\no \\ &&
 -\ep \sum _{k=\ma'+1}^\infty { \Gamma (k-\ep) \over \Gamma (1-\ep) \, k!}
{\Gamma (1+\ep)^2 \, \Gamma (k+a+1+3\ep) \Gamma (k+1+\ep)
\over \Gamma (k+a+2+4\ep) \Gamma (k+2+2\ep)}
\eea
where $\ma'=\max(0, -a-1)$. The summand behaves as $k^{ -3 -3\ep }$ for large $k$ and the infinite series is  absolutely and uniformly convergent in the neighborhood of $\ep=0$, and may be expanded in $\ep$.


\end{document}